\documentstyle[prd,aps,eqsecnum,preprint,tighten,floats,epsf]{revtex}
\begin{document}
\draft
\preprint{\vbox{\hfill UMN-D-98-2 \\
          \vbox{\hfill SLAC-PUB-7833}
          \vbox{\vskip0.3in}
          }}
\title{Nonperturbative renormalization \\
and the electron's anomalous moment
in large-$\alpha$ QED}

\author{John R. Hiller}
\address{%
Department of Physics, University of Minnesota--Duluth,
Duluth, Minnesota~~55812}

\author{Stanley J. Brodsky%
\footnote{\baselineskip=14pt
Work supported in part by the Department of Energy,
contract DE-AC03-76SF00515.}}
\address{%
Stanford Linear Accelerator Center,
Stanford University, Stanford, California 94309}

\date{\today}

\maketitle

\begin{abstract}

We study the physical electron in quantum electrodynamics expanded on the
light-cone Fock space in order to address two problems:  (1) the physics of
the electron's anomalous magnetic moment $a_e$ in nonperturbative QED, and
(2) the practical problems of ultraviolet regularization and
renormalization in truncated nonperturbative light-cone Hamiltonian
theory.  We present results for $a_e$ computed in a light-cone gauge Fock
space truncated to include one bare electron and at most two photons; i.e.,
up to two photons in flight.  The calculational scheme uses an invariant
mass cutoff, discretized light-cone quantization (DLCQ), a Tamm--Dancoff
truncation of the Fock space, and a photon mass regulator.  We introduce
new weighting methods which greatly improve convergence to the continuum
within DLCQ.  Nonperturbative renormalization of the coupling and electron
mass are carried out, and a limit on the magnitude of the effective
physical coupling strength is computed.  A large renormalized coupling
strength $\alpha_R= 0.1$ is then used to make the nonperturbative effects
in the electron anomalous moment from the one-electron, two-photon Fock
state sector numerically detectable.

\end{abstract}

\pacs{11.15.Tk,11.10.Gh,12.20.Ds,02.60.Nm
\begin{center}(Submitted to Physical Review D.)\end{center}}
 
\narrowtext
 
\section{Introduction}

Many years ago Feynman issued the following challenge\cite{Feynman}:
``It seems that very little physical intuition has yet been developed
in this subject [of quantum electrodynamics].  In nearly every case
we are reduced to computing exactly the coefficient of some specific
term.  We have no way to get a general idea of the result to be
expected. \ldots\ \ As a specific challenge, is there any method
of computing the anomalous moment of the electron which, on first
rough approximation, gives a rough approximation to the $\alpha$
term and a crude one to $\alpha^2$; and when improved, increases
the accuracy of the $\alpha^2$ term, yielding a rough estimate
to $\alpha^3$ and beyond?''  This challenge was taken up by
Drell and Pagels\cite{DrellPagels}, who used a sideways dispersion
relation and low-energy theorems for Compton scattering\cite{LowECompton}
to construct consistency conditions for the anomalous moment.
Their approach did meet with some success, particularly in
understanding the sign of the $\alpha$ term; however, the
dispersion relation requires an ultraviolet cutoff, and
low-energy approximations of the integrand are not completely adequate.

We propose to meet Feynman's challenge by using discretized
light-cone\cite{Dirac} quantization\cite{LCQ} (DLCQ)\cite{EarlyDLCQ}.
By constructing the dressed electron state in Fock space we can
in principle compute physical properties of the electron
exactly\cite{Compton,Zakopane}.
In practice, various truncations are required, but the approach
remains nonperturbative.  Instead of producing an expansion in $\alpha$,
we produce an expansion in Fock particles,
{\em i.e}.\ the number of photons in flight.  The computation is equivalent
to a selective summation of graphs to all orders, but is actually
done by diagonalizing a matrix approximation of the light-cone
mass-squared operator.  In this form the
calculation becomes a testing ground for techniques of
nonperturbative renormalization.

\begin{figure}
\centerline{\epsfxsize=\columnwidth \epsfbox{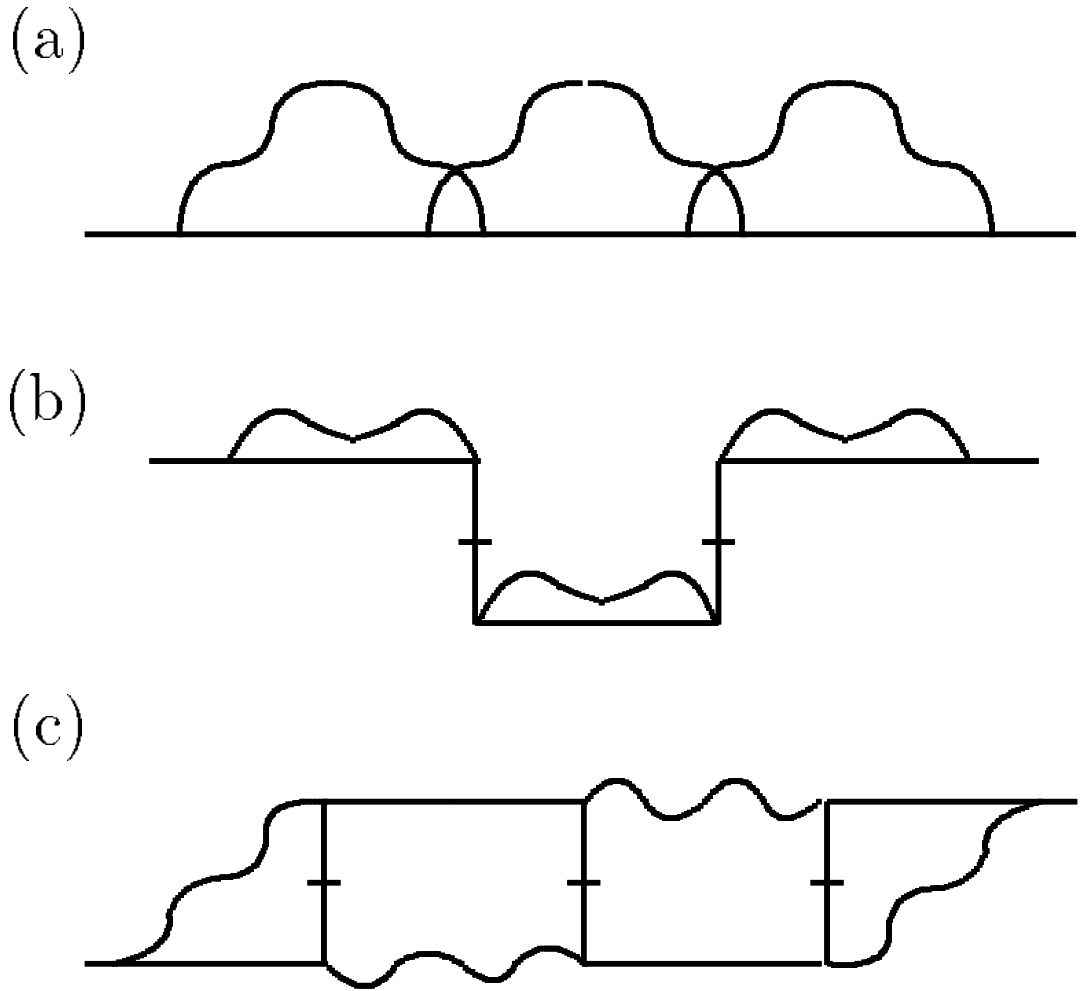} }
\caption{\label{fig:SampleGraphs}
Typical loop graphs that contribute to an infinite-order resummation.
Wavy lines represent photons and straight lines represent electrons.
A crossed line corresponds to an instantaneous exchange on the light cone.}
\end{figure}

When two-photon intermediate states are allowed, graphs such as the
multiloop graph in Fig.~\ref{fig:SampleGraphs}(a) enter the calculation and
are summed to all orders.  Even at the one-photon level, the
calculation is nonperturbative because there are infinite-order contributions
from graphs of the sort in Fig.~\ref{fig:SampleGraphs}(b) and (c).  However,
in the case of Fig.~\ref{fig:SampleGraphs}(c), crossed-photon graphs and
Z graphs are needed to cancel a divergence at zero longitudinal
momentum for any instantaneous electron.  Because of these
cancellations, we place the nonperturbative part of the one-photon
contribution into a two-photon calculation.

The selection of the graphs to be summed is driven by the truncations
made and is designed to make the calculation tractable.  The
truncation in particle number is physically reasonable; the work
of Drell and Pagels\cite{DrellPagels} shows that states with
few photons in flight are dominant. As the actual number of
particles is varied and as different truncations of the interactions
are explored, one can gain a better understanding of the physics
of the anomalous moment.  Thus our calculation can be viewed
as the beginning of a possible program, with systematic improvements
available.  It might even become a model for how to proceed with
nonperturbative calculations in quantum chromodynamics (QCD).

In the work presented here we regulate the theory by an invariant-mass cutoff,
which limits the light-cone energy of the Fock states included in
the basis, and by a Tamm-Dancoff truncation\cite{TammDancoff} of
the number of constituents.  These restrictions keep the numerical
calculation to a very reasonable size but they do complicate the
renormalization.  The truncations turn the bare parameters of
the theory into Fock-sector dependent functions of
momentum\cite{SectorDependent} and require careful construction of appropriate
counterterms\cite{Perry} to correctly approximate the solution to
the original theory\cite{ModelH}.  We fix these functions by applying
conditions on mass eigenvalues and vertices in the presence
of spectator constituents.

Two alternative renormalization procedures now exist.  One is the
similarity transformation developed by Wilson and
G{\l}azek\cite{SimTransform}  and Wegner\cite{Wegner},
where counterterms are generated perturbatively
as the Hamiltonian matrix is narrowed in the range of allowed
light-cone energies; the final Hamiltonian matrix is used
nonperturbatively. This approach has been applied by
Perry and co-workers\cite{PerryCoworkers}.
The other is based on introduction of
Pauli--Villars regulators\cite{PauliVillars} before the
quantization and numerical schemes are selected, so that
counterterms can be supplied simply by adjusting the bare
parameters.  This approach has just recently been successfully
tested in a simple model\cite{PV}, and it should soon be
considered for the anomalous-moment problem studied here.

We compute the anomalous moment\cite{Schwinger,Langnau} from a
spin-flip matrix element of the plus component of the
electromagnetic current\cite{BrodskyDrell}.
We approximate the Fock-state expansion of the dressed electron
with a truncation to no more
than two photons and one electron.
The eigenvalue problem for the wave functions
becomes a coupled set of
three integral equations.  To construct these equations
we use the light-cone Hamiltonian derived by Tang {\em et al}.\cite{Tang},
regulated by the invariant-mass cutoff.
The photon mass is taken to be one tenth of the electron mass,
to help control infrared divergences.
The coupling strength is set at $\alpha=1/10$, because
limitations on numerical accuracy make nonperturbative
effects discernible only at large coupling.  The calculation
is not an attempt to compete with the accuracy of the perturbative
calculations by Kinoshita and co-workers\cite{aePert}.

The bare electron mass in the one-photon sector is computed
from the one-loop correction allowed by the two-photon states,
where one photon is a spectator.
We then require that the bare mass in the no-photon sector
be such that physical mass is an eigenvalue of the light-cone
Hamiltonian.

The three-point bare coupling is also sector dependent.
There are no vacuum polarization effects, because pair production
is removed by the Tamm--Dancoff truncation. However, the
truncation violates the Ward identity so that vertex and
wave function renormalization do not cancel\cite{MustakiEtAl}.
A consequence of this is that the physical coupling is limited
to a cutoff-dependent finite range of allowed values.  We
compute the critical coupling that defines the upper limit.

The vertex renormalization is fixed by considering the proper part
of the transition amplitude for photon absorption by an electron
at zero photon momentum.  A means by which this transition
amplitude can be computed from the lowest eigenstate is
constructed.  Full diagonalization of the Hamiltonian is
not required; however, the renormalization condition and
the eigenvalue problem must be solved as a coupled system.
The wave function renormalization
is directly available from the bare amplitude in the Fock state
expansion.

To within the accuracy of the calculation, the values
for the anomalous moment become constant for an
ultraviolet cutoff sufficiently large.
However, most four-point graphs that arise in the bound-state problem
are log divergent.  To any order the divergences cancel if all graphs 
are included, but the Tamm--Dancoff truncation spoils this.
The resulting logarithmic effects are not detectable in the numerical
results.

The calculations presented here build on the
significant amount of work that has been done recently on
the use of light-cone quantization\cite{Dirac,LCQ,NearLC,Mtheory} in the
construction of solvable bound-state problems for strongly interacting
theories.  The coordinates used are based on the choice of $t+z$ as the time
coordinate, where $t$ is the ordinary time and $z$ any Cartesian
spatial coordinate.  A variety of field theories have been
analyzed in this way\cite{EarlyDLCQ,Vary,Eller,Hornbostel,Ma,%
Yukawa1D,Demeterfi,WC1+1,GrossNeveu,Sawicki,Wivoda,%
Yukawa,Tang,PauliPositronium,Hollenberg,PauliBayer,%
LadderRelations,TubeModel}.
Most theories considered have been simple model theories in one
space dimension\cite{EarlyDLCQ,Vary,Eller,Hornbostel,Ma,%
Yukawa1D,Demeterfi,WC1+1,GrossNeveu,%
LadderRelations,TubeModel}; however, there have been
studies of three dimensional theories, including the
Wick--Cutkosky model\cite{WickC,Sawicki,Wivoda}, the
Yukawa model\cite{Yukawa},
quantum electrodynamics (QED)\cite{Tang,PauliPositronium},
and quantum chromodynamics (QCD)\cite{Hollenberg}.
There has also been some work on nonperturbative scattering
calculations\cite{Kroger,JiSurya,Ree}
and on the stationary phase approximation
to the soliton in $\phi_{1+1}^4$\cite{KinkMass}.

Much of this work has involved numerical studies.
Brodsky, Pauli, and co-workers have analyzed various
one-dimensional theories (Yukawa\cite{EarlyDLCQ}, QED\cite{Eller}, and
QCD\cite{Hornbostel}) and have devoted a considerable amount
of effort to three-dimensional QED\cite{Tang,PauliPositronium}.  Some work
on the three-dimensional Wick--Cutkosky model\cite{WickC} has been done by
Sawicki and co-workers, Ji and Furnstahl\cite{Sawicki},
and Wivoda and Hiller\cite{Wivoda}.
One-dimensional scalar
theories, $\phi^3$ and $\phi^4$, have been studied by Harindranath
and Vary\cite{Vary}.  Work on the one-dimensional
Yukawa model has been done by Harindranath, Shigemitsu and
Perry\cite{Yukawa1D}; this was based on a Tamm--Dancoff
truncation\cite{TammDancoff} and used basis-function methods as well as a
discretization technique.  The basis-function methods have been extended
to the three-dimensional case by G{\l}azek {\em et al}.\cite{Yukawa}.
A preliminary treatment of QCD in three dimensions has been attempted by
Hollenberg and co-workers\cite{Hollenberg}.  Dimensional reduction of
QCD to an effective theory in $1+1$ dimensions has also been
considered\cite{LadderRelations,TubeModel}.

Other nonperturbative approaches applicable to QCD
include lattice gauge theory\cite{lattice,LatticeRenorm},
sum rules\cite{SumRules}, and Schwinger--Dyson
equations\cite{SchwingerDyson}.
A particularly successful form of lattice theory has been developed
by Lepage and collaborators\cite{Lepage} who use tadpole-improved
actions to reduce discretization errors.  The Hamiltonian form
of lattice theory\cite{KogutSusskind} is actually similar to the approach
usually taken in light-cone quantization, in that a Hamiltonian operator
is constructed and partially diagonalized\cite{Bronzan}.
There has also been work on combinations of the lattice with light-cone
methods in the transverse lattice method\cite{TransverseLattice} and
in direct use of a light-cone lattice\cite{lightconeLattice}.

Two important aspects of QCD that all these methods address are
vacuum structure and symmetry breaking.  In light-cone quantization, the
vacuum appears to be the trivial perturbative vacuum.  This has the advantage
that one can compute massive states immediately without first computing
the vacuum state.  However, in equal-time quantization, the nontrivial
structure of the QCD vacuum is known to be important.  This paradox of the
trivial vacuum has received much attention.  Nonperturbative analyses of
various light-cone models indicate that interactions induced by zero
modes\cite{KalloniatisRobertson,QEDzeromodes,ZeroModes} and
other considerations\cite{Perry} play important roles in generating effects
such as symmetry breaking\cite{SSB,Phi4} that are usually associated with the
vacuum.

The progress made recently in the application of light-cone quantization
owes much to earlier work\cite{EarlyWork}.  The development then was
aimed at perturbation theory, and in particular its application to
deep inelastic scattering, but much has been carried over to bound-state
problems. New work on perturbation theory has also been
done\cite{RecentPert,Langnau}.

An outline of the remainder of the paper is as follows.
The discretized light-cone formulation of the anomalous
moment problem is given in Sec.~\ref{sec:DLCQ}\@.
The nonperturbative mass and coupling
renormalization that we use are described in
Sec.~\ref{sec:Renormalization}.
Finite corrections associated with
photon zero modes and with ambiguities in infinite
renormalizations are discussed in Sec.~\ref{sec:FiniteCorrections}\@.
Our results are presented in Sec.~\ref{sec:Results},
and a brief summary is given in Sec.~\ref{sec:Summary}\@.

\section{Discretized light-cone quantization} \label{sec:DLCQ}

\subsection{Light-cone quantization}

We define light-cone coordinates\cite{LCQ} by
\begin{equation} \label{eq:coordinates}
x^\pm=t\pm z\,,\;\;\bbox{x}_\perp=(x,y)\,.
\end{equation}
Momentum variables are similarly constructed as
\begin{equation} \label{eq:momentum}
p^\pm=E\pm p_z\,,\;\;\bbox{p}_\perp=(p_x,p_y).
\end{equation}
The time variable is taken to be $x^+$, and time evolution of a system
is then determined by ${\cal P}^-$, the operator associated with the
momentum component conjugate to $x^+$.  Usually one seeks stationary
states obtained as eigenstates of ${\cal P}^-$.  Frequently the eigenvalue
problem is expressed in terms of a light-cone Hamiltonian\cite{EarlyDLCQ}
(mass-squared operator)
\begin{equation} \label{eq:HLC}
H_{\rm LC}={\cal P}^+{\cal P}^- - {\cal P}^2_\perp
\end{equation}
as
\begin{equation} \label{eq:EigenProb}
H_{\rm LC}\Psi=M^2\Psi\,,
\end{equation}
where $M$ is the mass of the state, and ${\cal P}^+$
and $\bbox{\cal P}_\perp$ are momentum operators
conjugate to $x^-$ and $\bbox{x}_\perp$.

It is convenient to work in a Fock basis
$\{|n:p_i^+,\bbox{p}_{\perp i}\rangle\}$ where ${\cal P}^+$
and $\bbox{\cal P}_\perp$ are diagonal, with $n$ the number
of particles and $i$ ranging between 1 and $n$.
To simplify the notation
only one particle type is included explicitly.
The state $\Psi$ is given by an expansion
\begin{equation} \label{eq:Psi}
\Psi=\sum_n\int [dx]_n\,[d^2k_\perp]_n\,
       \psi_n(x,\bbox{k}_\perp)
       |n:p^+=xP^+,\bbox{p}_\perp=x\bbox{P}_\perp+\bbox{k}_\perp\rangle\,,
\end{equation}
with
\begin{equation} \label{eq:dx}
[dx]_n=4\pi\delta(1-\sum_{i=1}^nx_i)
	     \prod_{i=1}^n\frac{dx_i}{4\pi\sqrt{x_i}}\,,\;\;\; 
[d^2k_\perp]_n=4\pi^2\delta(\sum_{i=1}^n\bbox{k}_{\perp i})
	       \prod_{i=1}^n\frac{d^2k_{\perp i}}{4\pi^2}\,,
\end{equation}
$(P^+,\bbox{P}_\perp)$ the total light-cone momentum, and $\psi_n$
interpreted as the wave function of the contribution from states
with $n$ particles.  The solution of (\ref{eq:EigenProb})
in principle yields these wave functions.

The anomalous moment $a_e$ is computed from the standard form factor
$F_2(q^2)$ at zero momentum transfer:
\begin{equation} \label{eq:aeDef}
a_e=F_2(0)\,.
\end{equation}
In the standard light-cone frame\cite{DrellYan} where
\begin{equation} \label{eq:lcframe}
q=(0,q_\perp^2/p^+,\bbox{q}_\perp=q_1\hat{x})\,,
\end{equation}
the form factor can be computed from the spin-flip matrix element of the plus
component of the current:
\begin{equation}  \label{eq:current}
-\frac{q_1}{2m_e}F_2(q^2)=
    \frac{1}{2p^+}\langle p+q,\uparrow|J^+(0)|p,\downarrow\rangle\,.
\end{equation}
Brodsky and Drell\cite{BrodskyDrell} have given a reduction of
this matrix element to a convenient form that depends directly 
on the wave functions.  From this we have
\begin{equation}  \label{eq:aeLC}
a_e=-2m_e\sum_je_j\sum_n\int\,[dx]_n\,[d^2k_\perp]_n\,
    \psi_{n\uparrow}^*(x,\bbox{k}_\perp)
    \sum_{i\neq j}x_i\frac{\partial}{\partial k_{1i}}
    \psi_{n\downarrow}(x,\bbox{k}_\perp)\,,
\end{equation}
where $e_j$ is the fractional charge of the struck constituent.

Up to this point, we have used formally exact expressions.
A key approximation to be
made is the truncation of all sums to a finite number of particles.
The result is the light-cone equivalent
of the Tamm-Dancoff approximation\cite{TammDancoff}.  The eigenvalue
problem becomes a finite set of equations that are in principle solvable.
However, the truncation has many consequences for the renormalization
of the theory\cite{SectorDependent} and for comparisons to Feynman
perturbation theory\cite{RecentPert,Langnau}. Some of these consequences
are discussed in Sec.~\ref{sec:Renormalization}.

In addition, QED requires regularization and renormalization.
To regularize it,
we use a cutoff on the invariant mass of the allowed Fock
states\cite{LCQ}
\begin{equation} \label{eq:cutoff}
\sum_i \frac{m_i^2+k_{\perp i}^2}{x_i}\leq\Lambda^2\,.
\end{equation}
This limits the relative transverse momentum $k_\perp$ of each
constituent and keeps the longitudinal momentum away from zero.
The latter aspect is important for control of spurious infrared
singularities, which are discussed in \ref{sec:Infrared}.
An additional cutoff, that limits the change in invariant mass
across any matrix element of the Hamiltonian\cite{LocalCutoff},
could be considered.

\begin{figure}
\centerline{\epsfxsize=\columnwidth \epsfbox{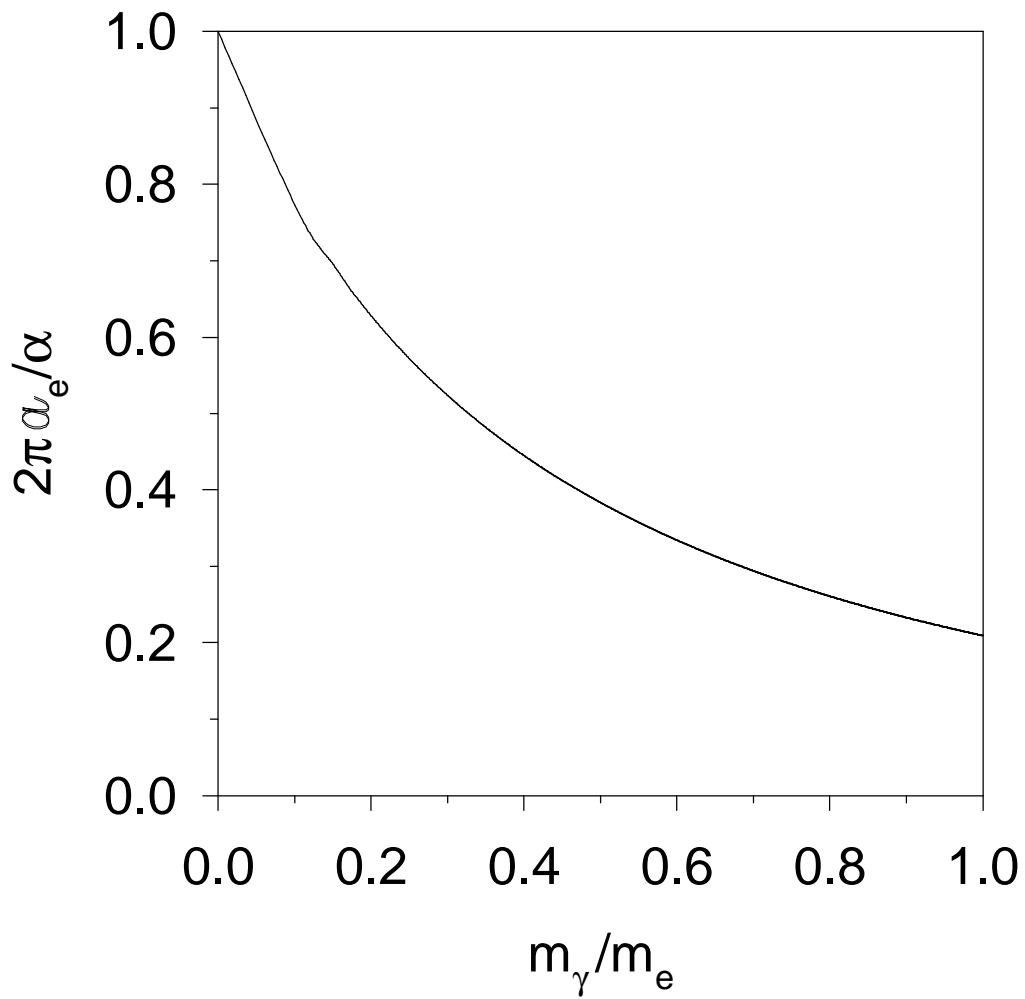} }
\caption{\label{fig:aeVSmgamma}
The one-photon perturbative contribution to the anomalous moment $a_e$ as
a function of photon mass $m_\gamma$.  It is the Schwinger term, given
by Eq.~\ref{eq:aeOnePhoton} of the text.}
\end{figure}

When only states with at most one photon and no pairs are retained,
and instantaneous interactions are neglected,
Brodsky and Drell\cite{BrodskyDrell} have shown that Eq.~(\ref{eq:aeLC})
reduces to
\begin{equation} \label{eq:aeIntegrals}
a_e=\frac{\alpha m_e^2}{\pi^2}\int\,dx\,d^2k_\perp\,\frac{m_e}{1-x}
   \frac{\theta(\Lambda^2-(m_e^2+k_\perp^2)/(1-x)-(m_\gamma^2+k_\perp^2)/x)}
          {[m_e^2-(m_e^2+k_\perp^2)/(1-x)-(m_\gamma^2+k_\perp^2)/x]^2}\,,
\end{equation}
which in the limit of $\Lambda\rightarrow\infty$ becomes
\begin{equation} \label{eq:aeOnePhoton}
a_e=a_e^S\equiv
  \frac{\alpha}{2\pi}\int_0^1\frac{2x^2(1-x)dx}{x^2+(1-x)(m_\gamma/m_e)^2}\,.
\end{equation}
Because the instantaneous interactions are higher order in $\alpha$,
this is the leading perturbative result.
The integrals involved can all be done analytically even for finite cutoff,
although the final
form is not instructive.  For $m_\gamma=0$, the resulting formula
yields the standard Schwinger\cite{Schwinger} contribution of $\alpha/2\pi$
at infinite cutoff.
In general this provides a point of comparison
for numerical calculations with one or more photons.  The inclusion of
the dependence on the photon mass in the analytic result is crucial
for comparison with numerical results calculated with nonzero $m_\gamma$
because the mass dependence is
quite strong, as can be seen in Fig.~\ref{fig:aeVSmgamma}.

\subsection{Discretization}

The most systematic approach to discretization of the eigenvalue problem
is the method originally suggested by Pauli and
Brodsky\cite{EarlyDLCQ}, discretized light-cone quantization (DLCQ).  In
essence it is the replacement of integrals by trapezoidal approximations,
with equally-spaced intervals in the longitudinal and transverse momenta
\begin{equation} \label{eq:DiscreteMom}
p^+\rightarrow\frac{\pi}{L}n\,,\;\;
\bbox{p}_\perp\rightarrow(\frac{\pi}{L_\perp}n_x,\frac{\pi}{L_\perp}n_y)\,.
\end{equation}
The length scales $L$ and $L_\perp$ determine the resolution of the
calculation.  Because the plus component of momentum is always
positive, the limit $L\rightarrow\infty$ can be exchanged for a limit
in terms of the integer {\em resolution}\cite{EarlyDLCQ}
\begin{equation} \label{eq:resolution}
K\equiv\frac{L}{\pi}P^+\,.
\end{equation}
The combination of momentum components that defines $H_{\rm LC}$ is then
independent of $L$.  The longitudinal momentum fractions $x_i$ become
ratios of integers $n_i/K$.  Because the $n_i$ are all positive, DLCQ
automatically limits the number of particles to no more than $K$.
The integers $n_x$ and $n_y$ range between limits associated with
some maximum integer $N_\perp$ fixed by the invariant-mass cutoff.
A finite matrix problem is then obtained without an explicit
Tamm-Dancoff truncation; however, this number of particles is much
too large in practice for numerical treatments of three-dimensional theories.

We use antiperiodic boundary conditions for the fermions and periodic
boundary conditions for the photons.  These restrict the integers $n$
associated with longitudinal momenta to being odd for fermions and even
for photons.  The description of the dressed electron state must then
use odd values of $K$.

In most applications, DLCQ is introduced at the level of second
quantization.  This can yield a compact expression of the eigenvalue
problem.  Recently, a transformation of the DLCQ Hamiltonian to a
Gaussian basis has been suggested\cite{Koures}; however, the
steps for renormalization in that basis have not been worked out.

The application of DLCQ to QED is summarized in \cite{Tang}, which we
use as a starting point.
This includes use of light-cone gauge \cite{WeylGauge}, with
$A^+=0$.  Modifications of this gauge choice
due to zero modes\cite{QEDzeromodes}
are discussed in Sec.~\ref{sec:FiniteCorrections}.

\begin{figure}
\centerline{\epsfxsize=\columnwidth \epsfbox{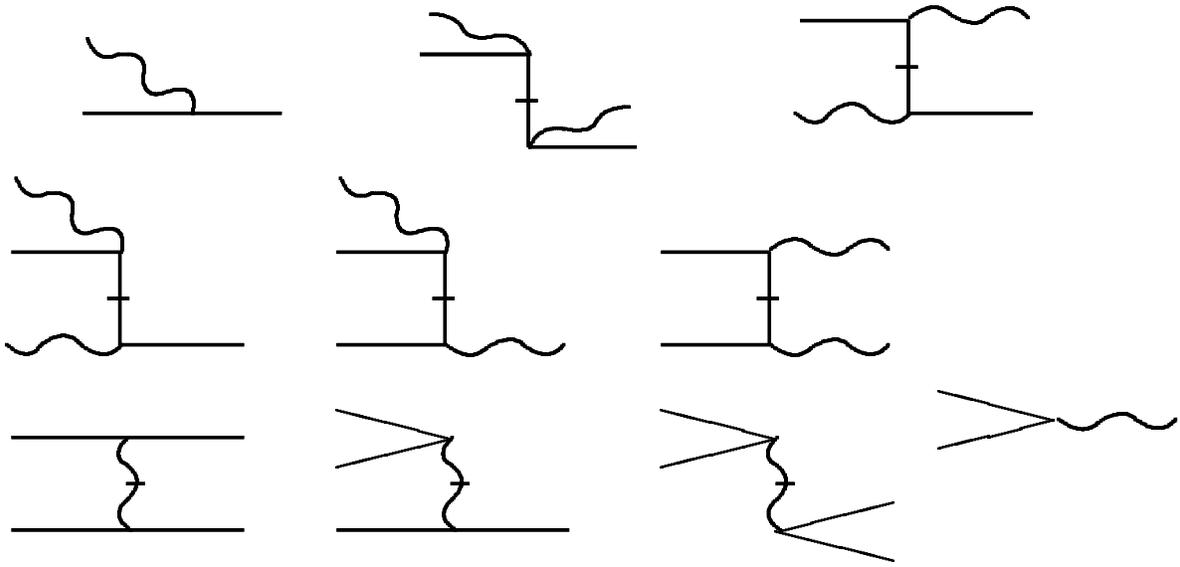} }
\caption{\label{fig:QEDgraphs}
Complete set of diagrams for the fundamental interactions of QED in
light-cone quantization.  Solid lines represent fermions; wavy
lines represent photons.  A line with a bar through it indicates
an instantaneous interaction expressed in the Hamiltonian as
a four-body operator.  Only the first three diagrams are included
in the present calculation, and for the third, the piece
kinematically equivalent to a Z graph is neglected.}
\end{figure}

The fundamental interactions of
light-cone QED are illustrated in Fig.~\ref{fig:QEDgraphs}.
For the calculation reported here we do not include any
pair production processes.  The instantaneous photon
interactions are then completely excluded because each
Fock state has only one fermion.  We also exclude the
fourth diagram (and its conjugate) to decouple two-photon
states from the bare electron state; this simplifies the
calculation and limits the role of two-photon states to
that of providing the basis for inclusion of crossed-photon
graphs.  There is also a technical modification of
the interaction associated with the third diagram of
Fig.~\ref{fig:QEDgraphs} which is discussed in Sec.~\ref{sec:Infrared}.
After the exclusions have been made, the light-cone Hamiltonian
becomes
\begin{eqnarray}
\lefteqn{H_{\rm LC}=\sum_{\underline{n}} \sum_{s=\pm 1/2}
              \frac{m_e^2+(\bbox{n}_\perp\pi/L_\perp)^2}{n/K}
                 b_{\underline{n},s}^\dagger b_{\underline{n},s}
   +\sum_{\underline{m}}  \sum_{\lambda=\pm 1}
              \frac{m_e^2+(\bbox{m}_\perp\pi/L_\perp)^2}{m/K}
                 a_{\underline{m},\lambda}^\dagger a_{\underline{m},\lambda}}
  \hspace{0.5in} \\
&&+\frac{eKm_e}{2\sqrt{\pi}L_\perp}
       \sum_{\underline{n}_1,\underline{n}_2,\underline{m}} \sum_{s=\pm 1/2}
       \frac{1}{\sqrt{m}}
           \left\{\delta_{\underline{n}_1+\underline{m},\underline{n}_2}
                \left(\frac{1}{n_1}-\frac{1}{n_2}\right)
                   b_{\underline{n}_2,s}^\dagger b_{\underline{n}_1,-s}
                        a_{\underline{m},2s} +\,\mbox{h.c.}\right\}
\nonumber \\
& & +\frac{e\sqrt{\pi}K}{\sqrt{2}L_\perp^2}
       \sum_{\underline{n}_1,\underline{n}_2,\underline{m}} \sum_{s=\pm 1/2}
     \frac{1}{\sqrt{m}} \delta_{\underline{n}_1+\underline{m},\underline{n}_2}
               \left\{ \bbox{\epsilon}_{\perp\,, 2s}\cdot
           \left(\frac{\bbox{m}_\perp}{m}-\frac{\bbox{n}_{\perp 1}}{n_1}\right)
                   b_{\underline{n}_2,s}^\dagger b_{\underline{n}_1,s}
                        a_{\underline{m},2s} +\,\mbox{h.c.}\right.
\nonumber \\
& & \hspace{1.5in} +\left. \bbox{\epsilon}_{\perp\,, -2s}\cdot
          \left(\frac{\bbox{m}_\perp}{m}-\frac{\bbox{n}_{\perp 2}}{n_2}\right)
                   b_{\underline{n}_2,s}^\dagger b_{\underline{n}_1,s}
                        a_{\underline{m},-2s} +\,\mbox{h.c.}\right\}
\nonumber \\
& &+\frac{e^2 K}{4\pi L_\perp^2}
\sum_{\underline{n}_1,\underline{n}_2,\underline{m}_1,\underline{m}_2}
    \sum_{s=\pm 1/2}   \frac{1}{\sqrt{m_1 m_2}}
      \delta_{\underline{n}_2+\underline{m}_2,\underline{n}_1+\underline{m}_1}
   \left\{\frac{1}{n_1+m_1} b_{\underline{n}_2,s}^\dagger b_{\underline{n}_1,s}
            a_{\underline{m}_2,-2s}^\dagger a_{\underline{m}_1,-2s}\right.
\nonumber \\
& & \hspace{1.5in} + \left. \frac{1}{m_2-n_1}
             b_{\underline{n}_2,s}^\dagger b_{\underline{n}_1,s}
              a_{\underline{m}_2,2s}^\dagger a_{\underline{m}_1,2s}\right\}\,.
\nonumber
\end{eqnarray}
with $\bbox{\epsilon}_{\perp\,,\lambda}=-(\lambda,i)/\sqrt{2}$.

The discrete form of the spin-$s$ eigenstate $\Psi_s$ is
\begin{eqnarray}
\Psi_s&=&\psi_{0s} b_{\underline{K},s}^\dagger|0\rangle+
\sum_{\underline{n},\underline{m}}
    \delta_{\underline{n}+\underline{m},\underline{K}}
      \sum_{s_1,\lambda_1}\psi_{1s}(\underline{n},\underline{m};s_1,\lambda_1)
   b_{\underline{n},s_1}^\dagger a_{\underline{m},\lambda_1}^\dagger|0\rangle
   \nonumber \\
& &+\sum_{\underline{n},\underline{m}_1,\underline{m}_2}
    \delta_{\underline{n}+\underline{m}_1+\underline{m}_2,\underline{K}}
      \sum_{s_1,\lambda_1,\lambda_2}
  \psi_{2s}(\underline{n},\underline{m}_1,\underline{m}_2;
                                            s_1,\lambda_1,\lambda_2)
   \frac{1}{\sqrt{2}}
    b_{\underline{n},s_1}^\dagger a_{\underline{m}_1,\lambda_1}^\dagger
    a_{\underline{m}_2,\lambda_2}^\dagger|0\rangle\,,
\end{eqnarray}
where $\underline{K}=(K,\bbox{K}_\perp=0)$.  According to the eigenvalue
equation $H_{\rm LC}\Psi=M^2\Psi$ the amplitudes $\psi_{is}$ must
satisfy the following (discretized) integral equations:
\begin{eqnarray}  \label{eq:Psi0s}
\lefteqn{(M^2-m_0^2)\psi_{0s}
 =e_0\frac{Km_e}{2\sqrt{\pi}L_\perp}\sum_{\underline{n},\underline{m}}
       \delta_{\underline{n}+\underline{m},\underline{K}}\frac{1}{\sqrt{m}}
        \left(\frac{1}{n}-\frac{1}{K}\right)
         \psi_{1s}(\underline{n},\underline{m};-s,2s)  }
 \hspace{0.5in}  \\
&  &
+e_0\frac{K\sqrt{\pi}}{\sqrt{2}L_\perp^2}\sum_{\underline{n},\underline{m}}
       \delta_{\underline{n}+\underline{m},\underline{K}}\frac{1}{\sqrt{m}}
       \left\{\bbox{\epsilon}_{\perp\,, 2s}\cdot
        \left(\frac{\bbox{m}_\perp}{m}-\frac{\bbox{n}_\perp}{n}\right)
          \psi_{1s}(\underline{n},\underline{m};s,2s) \right.
\nonumber \\
& & \hspace{1.5in} +\left.\bbox{\epsilon}_{\perp\,, -2s}\cdot
    \frac{\bbox{m}_\perp}{m}\psi_{1s}(\underline{n},\underline{m};s,-2s)
       \right\}\,,
\nonumber
\end{eqnarray}
\begin{eqnarray}
\lefteqn{\left(M^2
-\left[\frac{m_1^2(\underline{n}')+(\bbox{n}'_\perp \pi/L_\perp)^2}{n'/K}+
       \frac{m_\gamma^2+(\bbox{m}'_\perp \pi/L_\perp)^2}{m'/K}\right]\right)
         \psi_{1s}(\underline{n}',\underline{m}';s_1,\lambda_1) }
\hspace{0.5in}  \nonumber \\
&  &
=\frac{K}{4\pi L_\perp^2} \sum_{\underline{n},\underline{m}}
  \delta_{\underline{n}+\underline{m},\underline{K}} \frac{1}{\sqrt{m'm}}
  \left\{e_0^2\delta_{\lambda_1,-2s_1}\frac{1}{K}
     \psi_{1s}(\underline{n},\underline{m};s_1,-2s_1) \right.
\nonumber \\
&  &  \hspace{1in}
\left.-e_1(\underline{n}')e_1(\underline{n})\delta_{\lambda_1,2s_1}
     \frac{\theta(n-m')}{n-m'}
     \psi_{1s}(\underline{n},\underline{m};s_1,2s_1)\right\}
\nonumber \\
&  &
+e_0\frac{Km_e}{2\sqrt{\pi}L_\perp}\frac{1}{\sqrt{m'}}
        \left(\frac{1}{n'}-\frac{1}{K}\right)
          \delta_{\lambda_1,2s}\delta_{s_1,-s}\psi_{0s}
 \label{eq:Psi1s} \\
& &
+e_0\frac{K\sqrt{\pi}}{\sqrt{2}L_\perp^2}\frac{1}{\sqrt{m'}}
   \left\{\bbox{\epsilon}_{\perp\,,\lambda_1}^*\cdot
        \left(\frac{\bbox{m}'_\perp}{m'}-\frac{\bbox{n}'_\perp}{n'}\right)
          \delta_{\lambda_1,2s}\delta_{s_1,s}
  +\bbox{\epsilon}_{\perp\,,\lambda_1}^*\cdot
    \frac{\bbox{m}'_\perp}{m'}\delta_{\lambda_1,-2s}\delta_{s_1,s}\right\}
       \psi_{0s}
\nonumber \\
& &
+e_1(\underline{n}')
   \frac{Km_e}{2\sqrt{\pi}L_\perp}\sum_{\underline{n},\underline{m}}
       \delta_{\underline{n}+\underline{m}+\underline{m'},\underline{K}}
       \frac{1}{\sqrt{2m}}\left(\frac{1}{n}-\frac{1}{n+m}\right)
\nonumber \\
& &   \hspace{1in} \times
\left\{\psi_{2s}(\underline{n},\underline{m},\underline{m}';-s_1,2s_1,\lambda_1)
 +\psi_{2s}(\underline{n},\underline{m}',\underline{m};
                                          -s_1,\lambda_1,2s_1)\right\}
\nonumber \\
& &
+e_1(\underline{n}')
   \frac{K\sqrt{\pi}}{\sqrt{2}L_\perp^2}\sum_{\underline{n},\underline{m}}
     \delta_{\underline{n}+\underline{m}+\underline{m'},\underline{K}}
   \frac{1}{\sqrt{2m}}\left\{
      \bbox{\epsilon}_{\perp\,, 2s_1}\cdot
      \left(\frac{\bbox{m}_\perp}{m}-\frac{\bbox{n}_\perp}{n}\right)
      \psi_{2s}(\underline{n},\underline{m},\underline{m}';s_1,2s_1,\lambda_1)
      \right.
\nonumber \\
& &  \hspace{1in}
   +\bbox{\epsilon}_{\perp\,, 2s_1}\cdot
      \left(\frac{\bbox{m}_\perp}{m}-\frac{\bbox{n}_\perp}{n}\right)
      \psi_{2s}(\underline{n},\underline{m}',\underline{m};s_1,\lambda_1,2s_1)
\nonumber \\
& &   \hspace{1in}
   +\bbox{\epsilon}_{\perp\,, -2s_1}\cdot
      \left(\frac{\bbox{m}_\perp}{m}-\frac{\bbox{n}'_\perp}{n'}\right)
    \psi_{2s}(\underline{n},\underline{m},\underline{m}';s_1,-2s_1,\lambda_1)
\nonumber \\
& &   \hspace{1in}
\left.   +\bbox{\epsilon}_{\perp\,, -2s_1}\cdot
      \left(\frac{\bbox{m}_\perp}{m}-\frac{\bbox{n}'_\perp}{n'}\right)
     \psi_{2s}(\underline{n},\underline{m}',\underline{m};s_1,\lambda_1,-2s_1)
      \right\}\,,
\nonumber
\end{eqnarray}
and
\begin{eqnarray}
\lefteqn{\left(M^2-\left[\frac{m_e^2+(\bbox{n}'_\perp \pi/L_\perp)^2}{n'/K}
       +\frac{m_\gamma^2+(\bbox{m}'_{\perp 1} \pi/L_\perp)^2}{m'_1/K}
       \right. \right.}
\hspace{0.5in}  \nonumber \\
& & \hspace{2in} \left. \left.
       +\frac{m_\gamma^2+(\bbox{m}'_{\perp 2} \pi/L_\perp)^2}{m'_2/K}
                                                         \right]\right)
          \psi_{2s}(\underline{n}',\underline{m}'_1,\underline{m}'_2;
                                           s_1,\lambda_1,\lambda_2)
\nonumber \\
& &
=\frac{Km_e}{2\sqrt{2\pi}L_\perp}
\left\{\delta_{\lambda_2,-2s_1}
       \frac{e_1(\underline{n}'+\underline{m}'_2)}{\sqrt{m'_2}}
        \left(\frac{1}{n'}-\frac{1}{n'+m'_2}\right)
    \psi_{1s}(\underline{n}'+\underline{m}'_2,\underline{m}'_1;-s_1,\lambda_1)
\right. \nonumber \\
& &   \hspace{0.5in}
\left. +\delta_{\lambda_1,-2s_1}
          \frac{e_1(\underline{n}'+\underline{m}'_1)}{\sqrt{m'_1}}
        \left(\frac{1}{n'}-\frac{1}{n'+m'_1}\right)
     \psi_{1s}(\underline{n}'+\underline{m}'_1,\underline{m}'_2;-s_1,\lambda_2)
          \right\}
\nonumber \\
& &
\frac{K\sqrt{\pi}}{2L_\perp^2}\left\{
    \delta_{\lambda_2,2s_1}
      \frac{e_1(\underline{n}'+\underline{m}'_2)}{\sqrt{m_2}}
      \bbox{\epsilon}_{\perp\,, \lambda_2}^*\cdot
      \left(\frac{\bbox{m}'_{\perp 2}}{m'_2}-\frac{\bbox{n}'_\perp}{n'}\right)
      \psi_{1s}(\underline{n}'+\underline{m}'_2,\underline{m}'_1;s_1,\lambda_1)
\right.  \label{eq:Psi2s}  \\
& &   \hspace{0.5in}
   + \delta_{\lambda_1,2s_1}
      \frac{e_1(\underline{n}'+\underline{m}'_1)}{\sqrt{m_1}}
      \bbox{\epsilon}_{\perp\,, \lambda_1}^*\cdot
      \left(\frac{\bbox{m}'_{\perp 1}}{m'_1}-\frac{\bbox{n}'_\perp}{n'}\right)
      \psi_{1s}(\underline{n}'+\underline{m}'_1,\underline{m}'_2;s_1,\lambda_2)
\nonumber \\
& &    \hspace{0.5in}
   + \delta_{\lambda_2,-2s_1}
      \frac{e_1(\underline{n}'+\underline{m}'_2)}{\sqrt{m_2}}
      \bbox{\epsilon}_{\perp\,, \lambda_2}^*\cdot
      \left(\frac{\bbox{m}'_{\perp 2}}{m'_2}
              -\frac{\bbox{n}'_\perp+\bbox{m}'_{\perp 2}}{n'+m'_2}\right)
      \psi_{1s}(\underline{n}'+\underline{m}'_2,\underline{m}'_1;s_1,\lambda_1)
 \nonumber \\
& &    \hspace{0.5in}
\left.    + \delta_{\lambda_1,-2s_1}
      \frac{e_1(\underline{n}'+\underline{m}'_1)}{\sqrt{m_1}}
      \bbox{\epsilon}_{\perp\,, \lambda_1}^*\cdot
      \left(\frac{\bbox{m}'_{\perp 1}}{m'_1}
      -\frac{\bbox{n}'_\perp+\bbox{m}'_{\perp 1}}{n'+m'_1}\right)
      \psi_{1s}(\underline{n}'+\underline{m}'_1,\underline{m}'_2;s_1,\lambda_2)
      \right\}\,.  \nonumber
\end{eqnarray}
In anticipation of the discussion of renormalization in
Sec.~\ref{sec:Renormalization}, bare masses $m_0$ and $m_1$ and
bare couplings $e_0$ and $e_1$ have been introduced.
These equations are solved numerically, with the first step being the
use of (\ref{eq:Psi2s}) to eliminate $\psi_2$ from (\ref{eq:Psi1s}).
Once a solution is obtained
for one value of total spin $s$, the solution for the opposite spin
can be computed directly from
\begin{equation}
\psi_{1\downarrow}(s_1=-1/2,\lambda_1=\mp 1)=-\psi_{1\uparrow}^*(+1/2,\pm 1)
\end{equation}
and
\begin{equation}
\psi_{1\downarrow}(s_1=+1/2,\lambda_1=\mp 1)=+\psi_{1\uparrow}^*(-1/2,\pm 1)\,.
\end{equation}
These follow from the symmetries of the integral equations.

Other symmetries lead to relationships between amplitude components,
which can be summarized as follows:
\begin{equation}
\psi_{1\uparrow}(x,\bbox{k}_\perp;s_1=1/2,\lambda_1=\pm 1)=
    k_xf_{r\lambda_1}(|k_x|,|k_y|)+ik_yf_{i\lambda_1}(|k_x|,|k_y|)
\end{equation}
and
\begin{equation}
\psi_{1\uparrow}(x,\bbox{k}_\perp;s_1=-1/2,\lambda_1=\pm 1)=
    g_{r\lambda_1}(|k_x|,|k_y|)+ik_xk_yg_{i\lambda_1}(|k_x|,|k_y|)\,,
\end{equation}
where the functions $f_{r\lambda}$, $f_{i\lambda}$, $g_{r\lambda}$,
and $g_{i\lambda}$ are real.  The problem
can then be reduced to a smaller matrix problem for these real functions.
For $s_1=1/2$ we store $k_xf_{r\pm}(k_x>0,0)$,
$k_yf_{i\pm}(0,k_y>0)$, $k_xf_{r\pm}(k_x>0,k_y>0)$, and
$k_yf_{i\pm}(k_x>0,k_y>0)$\@.
For $s_1=-1/2$ we store $g_{r\pm}(0,0)$, $g_{r\pm}(k_x>0,0)$,
$g_{r\pm}(0,k_x>0)$, $g_{r\pm}(k_x>0,k_y>0)$, and
$k_xk_yg_{i\pm}(k_x>0,k_y>0)$.
The use of symmetry reduces the matrix storage requirement by a factor of 8.
The Hermitian matrix of the original
eigenvalue equation (\ref{eq:Psi1s}) can be expressed as a real symmetric
matrix in the reduced equation by using a two-component representation
of complex arithmetic:
\begin{equation}
(c+id)(\alpha+i\beta)\longrightarrow
   \left(\begin{array}{rr} c & -d \\ d & c \end{array} \right)
   \left(\begin{array}{c} \alpha \\ \beta \end{array} \right)
\end{equation}
and
\begin{equation}
(\alpha+i\beta)^*\longrightarrow
   \left(\begin{array}{rr}1 & 0 \\ 0 & -1 \end{array} \right)
    \left(\begin{array}{c} \alpha \\ \beta \end{array} \right)\,.
\end{equation}

The leading perturbative result\cite{BrodskyDrell} is recovered
by keeping only $\psi_{0s}$ terms on the right-hand side of
(\ref{eq:Psi1s}).  This equation can then be immediately solved
for $\psi_{1s}$, which can be used to form a discrete approximation
to (\ref{eq:aeIntegrals}).  The approximation includes a finite difference
approximation to the derivatives that appear in (\ref{eq:aeLC}) and
therefore is not simply a trapezoidal approximation to (\ref{eq:aeIntegrals}).

\subsection{Discretization errors}

\begin{figure}
\centerline{\epsfxsize=\columnwidth \epsfbox{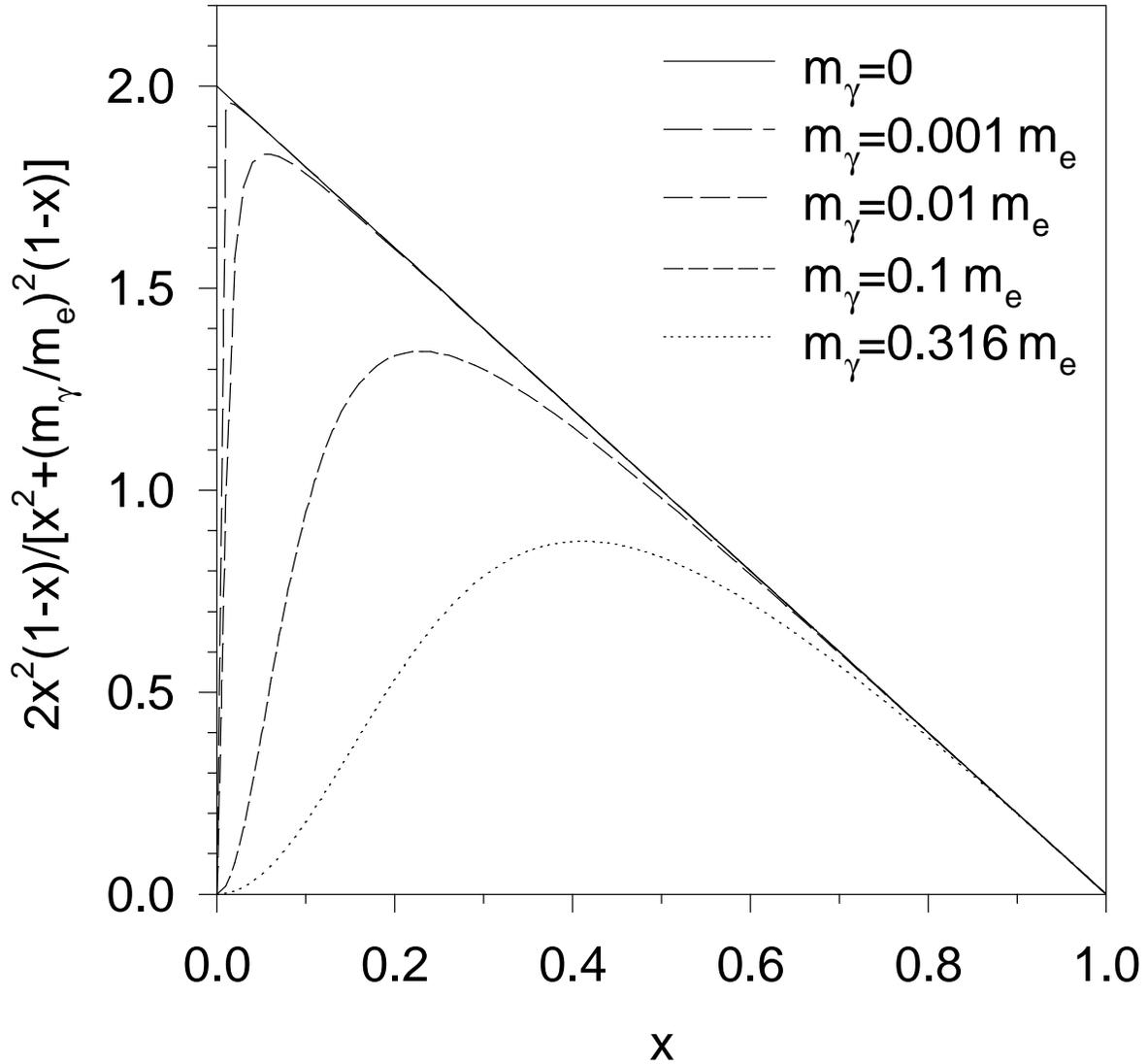} }
\caption{\label{fig:OnePhoton}
The integrand for the one-photon perturbative contribution.  The integrals
over the transverse momentum have been performed and only the
integral over longitudinal momentum $x$ remains, as given in
Eq.~(\protect\ref{eq:aeOnePhoton}).}
\end{figure}

Results obtained with ordinary DLCQ show an
irregular dependence on the numerical parameters $K$ and
$N_\perp$, which interferes with extrapolation to infinite resolution.
The causes of the irregularities have been determined to be the
numerical approximation of the derivative in the formula (\ref{eq:aeLC}) for
$a_e$ and boundary effects in the numerical
integrations. The error in the derivative can be controlled by choosing 
$N_\perp\geq 7$ and $K\geq 21$.  The
bound on $K$ is consistent with the resolution needed to resolve 
the one-photon peak in the $a_e$ integrand when
$m_\gamma=m_e/10$, which is the photon mass we use.
Smaller values of $m_\gamma$ shift the peak to smaller photon
momenta and would increase the lower bound
on $K$.  The shape of the integrand for various values of $m_\gamma$
is illustrated in Fig.~\ref{fig:OnePhoton}.  The mass sensitivity
of the numerical convergence rate is shown in Table~\ref{tab:OnePhoton}.

\begin{table}
\mediumtext
\caption{\label{tab:OnePhoton}
The DLCQ approximation to the perturbative one-photon integral in
Eq.~(\protect\ref{eq:aeOnePhoton}).  The values at different DLCQ
resolutions $K$ show that the convergence rate depends on the photon
mass $m_\gamma$.}
\begin{tabular}{rddddd}
 & \multicolumn{5}{c}{$m_\gamma/m_e$}  \\
\cline{2-6}   \\
$K$ & 0 & 0.001 & 0.01 & 0.1 & 0.3162 \\
\hline
   11 &  0.8182 & 0.8182 & 0.8173 & 0.7455 & 0.5041  \\
   21 &  0.9048 & 0.9047 & 0.9025 & 0.7708 & 0.5079  \\
   41 &  0.9512 & 0.9512 & 0.9461 & 0.7728 & 0.5090  \\
   81 &  0.9753 & 0.9752 & 0.9648 & 0.7731 & 0.5093  \\
  161 &  0.9876 & 0.9873 & 0.9699 & 0.7732 & 0.5094  \\
  321 &  0.9938 & 0.9933 & 0.9703 & 0.7732 & 0.5094  \\
  641 &  0.9969 & 0.9959 & 0.9703 & 0.7732 & 0.5094  \\
 1281 &  0.9984 & 0.9968 & 0.9703 & 0.7732 & 0.5094  \\
 2561 &  0.9992 & 0.9969 & 0.9703 & 0.7732 & 0.5094  \\
$\infty$ &  1.0000 & 0.9969 & 0.9703 & 0.7732 & 0.5094 \\
\end{tabular}
\narrowtext
\end{table}

The integration boundary effects are more difficult to control.  These
effects arise from use of the DLCQ grid which is incommensurate with the
integration domain.  At the boundaries, the trapezoidal rule misses
contributions beyond the last grid point; this error is not a smooth
function of the grid spacing.  To overcome this error, one can replace
the trapezoidal rule by open-closed Newton--Cotes formulas tailored
specifically to the boundary \cite{PV}.
Grid points near the boundary are then
associated with unequal integration weights.  The unequal weights
must be taken into account in normalization sums and symmetrization
of the Hamiltonian matrix, but this is easily done.  One can even
consider use of Simpson's rule, although this does not appear useful
in the anomalous moment calculation.
The improvement brought by these weighting methods can be dramatic,
as shown in Ref.~\cite{PV}.

\section{Renormalization} \label{sec:Renormalization}

\subsection{Mass renormalization}

Here we are interested
in ultraviolet divergences associated with large $k_\perp$.
Electron self-energy contributions, which are divergent, shift
the mass and, through wave function renormalization, change the
coupling.  These induce $\Lambda^2$ and $\log\Lambda$ dependencies
in the eigenvalues.  In the discrete truncated problem these effects
depend on the Fock sector considered\cite{SectorDependent}.  For example,
an electron in a Fock state for which a transition to a state with
more photons is not allowed, perhaps due to truncation in photon number,
will not experience any self-energy corrections.  If one additional
photon is allowed, but not instantaneous interactions, only single loops
can occur.  If two or more additional photons can appear,
then an infinite number of overlapping loops can contribute to
self-energy corrections, a truly nonperturbative situation.
In each case, the leading divergence is removed by introduction of
counterterms associated with bare masses that are sector dependent
and momentum dependent\cite{Bvds}.

To be specific, consider the case where there are at most two photons
and only one electron.  The Fock-state expansion can be written
schematically as
\begin{equation} \label{eq:SchematicExpansion}
\Psi=\psi_0|e\rangle+\bbox{\psi}_1|e\gamma\rangle
	      +\bbox{\psi}_2|e\gamma\gamma\rangle\,.
\end{equation}
Here $\bbox{\psi}_1$ and $\bbox{\psi}_2$ are column vectors that contain the
amplitudes for individual Fock states with one and two photons, respectively.
The eigenvalue problem (\ref{eq:EigenProb}) becomes a coupled set of
three integral equations (\ref{eq:Psi0s}), (\ref{eq:Psi1s}),
and (\ref{eq:Psi2s}), which we write more compactly as
\begin{eqnarray} \label{eq:IntegralEqns}
m_0^2\psi_0 + \bbox{b}_1^\dagger\cdot\bbox{\psi}_1
     + \bbox{b}_2^\dagger\cdot\bbox{\psi}_2  & = & M^2\psi_0\,, \nonumber \\
\bbox{b}_1\psi_0 + \bbox{A}_{11}\bbox{\psi}_1
     + \bbox{A}_{12}\bbox{\psi}_2 & = & M^2\bbox{\psi}_1\,,   \\
\bbox{b}_2\psi_0 + \bbox{A}_{12}^\dagger\bbox{\psi}_1
     + \bbox{A}_{22}\bbox{\psi}_2 & = & M^2\bbox{\psi}_2\,,   \nonumber
\end{eqnarray}
where $m_0$ is the bare electron mass and the vectors $\bbox{b}_i^\dagger$
and the tensors $\bbox{A}_{ij}$ are integral operators
obtained from $H_{\rm LC}$.  We now
require that $m_0$ be such that $M^2=m_e^2$ is an eigenvalue.  The
second and third equations can be solved for $\bbox{\psi}_1/\psi_0$
and $\bbox{\psi}_2/\psi_0$.  Then the first equation yields $m_0$.
Normalization of $\Psi$ fixes the value of $\psi_0$.

Suppose now that this two-photon problem is embedded in some larger
problem where one needs to know the bare mass of an electron in a
Fock state that can couple to Fock states obtained by adding at most two
photons.  The same set of equations can be applied, with all constituents
in the lowest Fock state, other than the electron, acting as spectators.
One need only replace $m_0^2$ in (\ref{eq:IntegralEqns}) by
$(m_0^2+k_\perp^2)/x$ and $M^2=m_e^2$ by
$M^2=(m_e^2+k_\perp^2)/x$, with $x$ and $\bbox{k}_\perp$ the longitudinal
momentum fraction and transverse momentum of the initial electron.
Notice that $m_0$ is now a function of $x$ and $\bbox{k}_\perp$.

This can be generalized to cases with more photons, and reduced to
the case of only one contributing photon.
Thus one obtains a mechanism for a sector-dependent, momentum-dependent
mass renormalization that is used from the top $n$-photon sector down
to the bare electron state $|e\rangle$.  The last step automatically
includes the solution of the full eigenvalue problem for the
dressed electron state.

For the one-photon case embedded in the two-photon problem we
have
\begin{eqnarray}
\frac{m_1^2+k_\perp^2}{x}\psi_{1s}(x,\bbox{k}_\perp)
  +\bbox{b}^\dagger(x,\bbox{k}_\perp)\cdot\bbox{\psi}_2 & = &
           \frac{m_e^2+k_\perp^2}{x}\psi_{1s}(x,\bbox{k}_\perp) \,, \\
\bbox{b}\,\psi_{1s}(x,\bbox{k}_\perp) +\bbox{A}\bbox{\psi}_2 & = &
      \frac{m_e^2+k_\perp^2}{x}\bbox{\psi}_{2s}\,.
\end{eqnarray}
The second photon is a spectator.  The coupling to $\bbox{\psi}_2$
then induces the one-loop self-energy correction with this spectator
present.  The explicit form is obtained from (\ref{eq:Psi1s})
and (\ref{eq:Psi2s}) with
$M^2=(m_e^2+k_\perp^2)/x+(m_\gamma^2+k_\perp^2)/(1-x)$ and
with any interaction involving the spectator dropped.  Eq.~\ref{eq:Psi2s}
can then be solved for $\bbox{\psi}_2$ and the result substituted
into the modified (\ref{eq:Psi1s}) to obtain.
\begin{eqnarray}
\lefteqn{\frac{m_e^2+(\bbox{n}'_\perp \pi/L_\perp)^2}{n'/K}
         \psi_{1s}(\underline{n}',\underline{m}';s_1,\lambda_1) =
         \frac{m_1^2(\underline{n}')+(\bbox{n}'_\perp \pi/L_\perp)^2}{n'/K}
         \psi_{1s}(\underline{n}',\underline{m}';s_1,\lambda_1)
\hspace{0.5in} } \\
&  &  \hspace{0.25in}
+e_1^2(\underline{n}')\frac{K^2 m_e^2}{4\pi L_\perp^2}
    \sum_{\underline{n},\underline{m}}
  \frac{\delta_{\underline{n}+\underline{m},\underline{n}'}/m
    \left(1/n'-1/n\right)^2
    \psi_{1s}(\underline{n}',\underline{m}';s_1,\lambda_1)}
     {\frac{K}{n'}\left[m_e^2+(\bbox{n}'_\perp \pi/L_\perp)^2\right]
       -\frac{K}{n}\left[m_e^2+(\bbox{n}_\perp \pi/L_\perp)^2\right]
       -\frac{K}{m}\left[m_\gamma^2+(\bbox{m}_\perp \pi/L_\perp)^2\right]}
\nonumber \\
&  &  \hspace{0.25in}
+e_1^2(\underline{n}')\frac{K^2 \pi}{4 L_\perp^4}
    \sum_{\underline{n},\underline{m}}
  \frac{\delta_{\underline{n}+\underline{m},\underline{n}'}/m
   \left\{\left(\bbox{m}_\perp/m-\bbox{n}_\perp/n\right)^2
          +\left(\bbox{m}_\perp/m-\bbox{n}'_\perp/n'\right)^2
           \right\}\psi_{1s}(\underline{n}',\underline{m}';s_1,\lambda_1)}
   {\frac{K}{n'}\left[m_e^2+(\bbox{n}'_\perp \pi/L_\perp)^2\right]
       -\frac{K}{n}\left[m_e^2+(\bbox{n}_\perp \pi/L_\perp)^2\right]
       -\frac{K}{m}\left[m_\gamma^2+(\bbox{m}_\perp \pi/L_\perp)^2\right]}\,.
\nonumber
\end{eqnarray}
The Kronecker deltas from helicity conservation have been used to
simplify the result, and only terms in which the second photon
is a spectator have been kept.  Rearrangement of the coefficient
of $\psi_{1s}$ yields
\begin{eqnarray}
\lefteqn{m_1^2(\underline{n}')=m_e^2
   -\frac{n'}{K}e_1^2(\underline{n}')\frac{K^2\pi}{4L_\perp^4}}  \\
& & \hspace{0.5in}
  \times  \sum_{\underline{n},\underline{m}}
        \frac{\delta_{\underline{n}+\underline{m},\underline{n}'}}{m}
  \frac{\left(m_e L_\perp/\pi\right)^2
                  \left(1/n'-1/n\right)^2
          +\left(\bbox{m}_\perp/m-\bbox{n}_\perp/n\right)^2
          +\left(\bbox{m}_\perp/m-\bbox{n}'_\perp/n'\right)^2}
   {\frac{K}{n'}\left[m_e^2+(\bbox{n}'_\perp \pi/L_\perp)^2\right]
       -\frac{K}{n}\left[m_e^2+(\bbox{n}_\perp \pi/L_\perp)^2\right]
       -\frac{K}{m}\left[m_\gamma^2+(\bbox{m}_\perp \pi/L_\perp)^2\right]}
\nonumber
\end{eqnarray}
as the one-loop mass.

If electron-positron pairs are included, the photon mass is
renormalized and must be treated in the analogous fashion.  In general,
the two mass renormalizations are coupled, and must be carried out
simultaneously.

All of the steps in mass renormalization depend on knowing all
couplings.  This information is actually not immediately available
because the couplings are to be renormalized.

\subsection{Coupling renormalization}

The bare coupling for the electron-photon three-point vertex depends
on the initial and final momenta of the electron and on the sectors
between which the coupling acts\cite{SectorDependent}.  The momentum
dependence is present because the amount of momentum available constrains
the extent to which higher order corrections can contribute.  Similarly,
the sector dependence makes itself felt when the number of additional
particles in higher-order corrections is restricted.

We fix these bare coupling functions by matching photon absorption
amplitudes to the fundamental three-point vertex.  The amplitudes are
computed from the numerical eigenfunction of the light-cone Hamiltonian.
Therefore, the coupling renormalization conditions and the mass eigenvalue
problem form a coupled set of equations that are solved iteratively.

\subsubsection{Renormalization conditions}

When vacuum polarization is absent, the bare coupling $e_0$ is related
to the physical coupling $e_R$ by
\begin{equation}  \label{eq:e0}
e_0(\underline{k}_i,\underline{k}_f)
    =\frac{Z_1(\underline{k}_f)e_R}
          {\sqrt{Z_{2i}(\underline{k}_i)Z_{2f}(\underline{k}_f)}}
\end{equation}
where $\underline{k}_i=(k_i^+,\bbox{k}_{\perp i})$ is the initial electron
momentum and $\underline{k}_f$ the final momentum.  The renormalization
functions $Z_1(\underline{k})$ and $Z_2(\underline{k})$ are
generalizations of the usual constants\cite{MustakiEtAl}.

The wave function renormalization function $Z_2$ is easily computed
since it is the probability of the bare electron Fock state in the
dressed electron state. In the earlier notation of
Eq.~(\ref{eq:SchematicExpansion}), we have
\begin{equation}  \label{eq:Z2}
Z_2(\underline{k})=|\psi_0|^2\,,
\end{equation}
where $\underline{k}$ is the light-cone momentum of the dressed electron.
The amplitude $\psi_0$ must be computed in a basis where only allowed
particles appear.  For example, if the vertex is the photon absorption
process,
$Z_{2i}$ must be computed with one less photon in the basis than in the
basis used for $Z_{2f}$.  From this example one can see that the
Tamm-Dancoff approximation has destroyed the usual Ward
identity.

The function $Z_1$ can be fixed by considering the transition amplitude
$T_{fi}$ for photon absorption by an electron at zero photon momentum.
The proper part of this amplitude, meaning that without self-energy
corrections to the legs, is required to be proportional to the elementary
three-point no-flip vertex $V_{fi}$ when
$\underline{q}=\underline{k}_f-\underline{k}_i\rightarrow 0$:
\begin{equation}  \label{eq:CouplingRenormCond}
T_{fi}^{\rm proper}=\frac{1}{Z_1(\underline{k}_f)}V_{fi}\,.
\end{equation}
In the limit, only $\underline{k}_f=\underline{k}_i$ dependence can
remain.\footnote{There are, of course, finite corrections that are not
properly represented here.  These are discussed in
Sec.~\ref{sec:FiniteCorrections}.}
Numerically the limit can be taken by using a photon with momentum
$(2P^+/K,\hat{q}_\perp \pi/L_\perp)$; in the DLCQ limit of
$K\rightarrow \infty$ and $L_\perp\rightarrow\infty$, this momentum becomes
zero. Of course, if this particular state does not satisfy the cutoff,
a state with slightly larger longitudinal momentum must be used instead.

The full transition amplitude can be computed from solutions to the
eigenvalue problem (\ref{eq:EigenProb}).  Let $H_0$ be the free light-cone
Hamiltonian with physical masses.  The eigenstates of $H_0$ are then the
asymptotic states of the electron and photon.  The transition is driven
by the interaction $V\equiv H_{\rm LC}-H_0$.  Define resolvents for the
free and full Hamiltonians as
\begin{equation}  \label{eq:Resolvents}
G^+=\frac{1}{s+i\epsilon-H_0}\;\;\;\mbox{and}\;\;\;
{\cal G}^+=\frac{1}{s+i\epsilon-H_{\rm LC}}\,,
\end{equation}
with $s$ the square of the center-of-mass energy.  The $T$ matrix can be
formally expressed in terms of these as
\begin{equation}  \label{eq:LippmannSchwinger}
G^+TG^+=G^+VG^++G^+VG^+TG^+={\cal G}^+VG^+\,.
\end{equation}
When sandwiched between the initial and final states, this yields
\begin{equation}  \label{eq:Sandwich}
\frac{1}{s-m_e^2}T_{fi}\frac{1}{s-s_i}=
   \sum_n\frac{\psi_{n0}}{s-M_n^2}\langle\Psi_n|V|i\rangle\frac{1}{s-s_i}\,,
\end{equation}
where the $|\Psi_n\rangle$ are eigenstates of $H_{\rm LC}$ with eigenvalues
$M_n^2$ and bare-electron amplitudes $\psi_{n0}$.  In the limit\footnote{This
limit neglects the small photon mass.} that $s$ becomes $m_e^2$, we obtain
\begin{equation}  \label{eq:TfiFormula}
T_{fi}=\psi_0\langle\Psi|V|i\rangle\,,
\end{equation}
in which $|\Psi\rangle$ is the dressed electron state and
$\psi_0=\sqrt{Z_{2f}(\underline{k}_f)}$.

The connection between $T_{fi}$ and $T_{fi}^{\rm proper}$ is
made by considering the matrix element of
$G^+TG^+=G^+V\sum_{n=0}^\infty\left(G^+V\right)^nG^+$.  We have
\begin{equation} \label{eq:TproperMatrixElt}
\langle f|G^+TG^+|i\rangle=
  \sum\langle f|G^+V\cdots VG^+|f\rangle
     \langle f|VG^+V\cdots G^+|i\rangle
      \langle i|G^+V\cdots VG^+|i\rangle\,.
\end{equation}
\begin{figure}
\centerline{\epsfxsize=\columnwidth \epsfbox{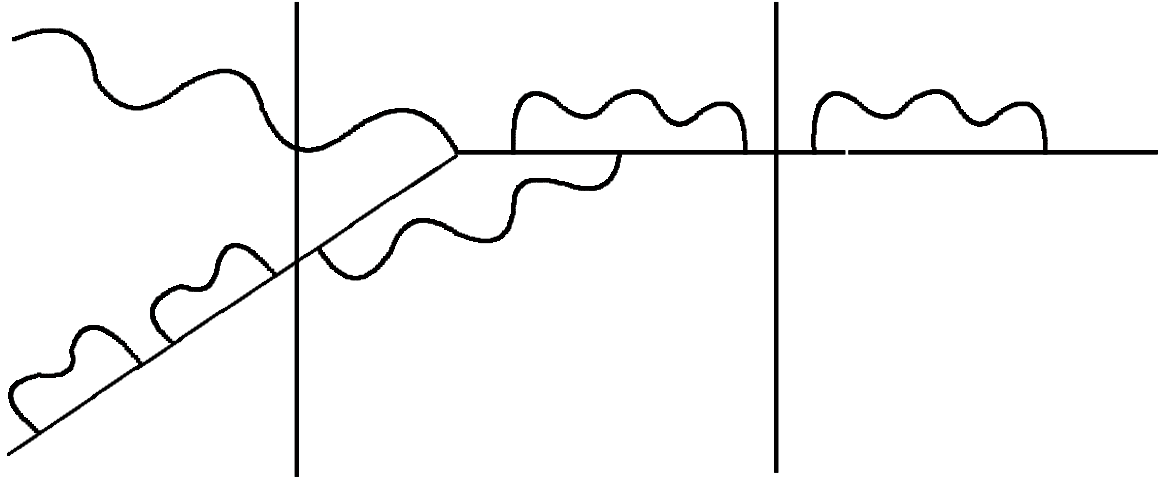} }
\caption{\label{fig:ProperVertex}
Representative diagram for the extraction of the proper
vertex amplitude $T_{fi}^{\rm proper}$ from the full amplitude $T_{fi}$.
The vertical lines separate the three regions of the diagram.  To the
left the initial photon is a spectator to the dressing of the initial
electron.  To the right only corrections
to the final electron line are present.  The proper vertex is in the
center.}
\end{figure}
The factors on the right are illustrated in Fig.~\ref{fig:ProperVertex}.
The second factor contains no intermediate $|f\rangle$ states
and the initial photon is absorbed before $|i\rangle$ appears
as an intermediate state.  In the third factor, the initial
photon remains a spectator throughout.  The sum runs over
all possible combinations of these forms and yields
\begin{equation} \label{eq:TproperSummed}
\langle f|G^+TG^+|i\rangle=\langle f|{\cal G}^+|f\rangle
   T_{fi}^{\rm proper} \langle e_i|{\cal G}_\gamma^+|e_i\rangle,
\end{equation}
where ${\cal G}_\gamma^+$ is the propagator for the
electron in the presence of the initial photon as a spectator.
In the limit where $s$ and $s_i$ approach $m_e^2$, we obtain
\begin{equation} \label{eq:Tproper}
\frac{1}{s-m_e^2}T_{fi}\frac{1}{s-s_i}=
        \frac{|\psi_{f0}|^2}{s-m_e^2}
                T_{fi}^{\rm proper}\frac{|\psi_{i0}|^2}{s-s_i}\,.
\end{equation}
This then reduces to an expression for $T_{fi}^{\rm proper}$
\begin{equation}  \label{eq:TproperFormula}
T_{fi}^{\rm proper}=\frac{1}{Z_{2f}Z_{2i}}T_{fi}\,.
\end{equation}
Thus the solution of the eigenvalue problem for only one state can be
used to compute $Z_1$.  Full diagonalization of $H_{\rm LC}$ is not
needed.

\subsubsection{Application of renormalization conditions}

Because $Z_1$ is needed in the construction of $H_{\rm LC}$,
the eigenvalue problem and the renormalization conditions
must be solved simultaneously.  This leads to an iterative procedure
that begins with an initial guess for the bare coupling functions.
One then computes bare masses and new bare couplings.  The process is
repeated until convergence is attained.  This must be done from the
top sector down; the bare masses in any one sector and the bare couplings
between any two depend only on the sectors above, the ones with more
photons.  The structure of the Hamiltonian matrix can then be determined
once and for all at these levels and then used in the determination
of the structure at the levels further below.

When the Fock basis is limited to no more than one photon,
and instantaneous interactions are neglected, the
renormalization conditions are quite simple.  We have from
(\ref{eq:TfiFormula})
\begin{equation}
T_{fi}=\psi_0\psi_0^*V_{fi}=Z_{2f}V_{fi}
\end{equation}
and from this, with (\ref{eq:CouplingRenormCond})
and (\ref{eq:TproperFormula}),
\begin{equation}
Z_1=Z_{2f}Z_{2i}\frac{V_{fi}}{Z_{2f}V_{fi}}=Z_{2i}=1\,,
\end{equation}
where the last equality follows from the unavailability of any
state that can correct the initial electron line when a photon
spectator is present.  The bare charge is then given by
\begin{equation} \label{eq:e1}
e_1=\frac{Z_1 e_R}{\sqrt{Z_{2f}Z_{2i}}}=\frac{e_R}{\psi_0}\,.
\end{equation}
The subscript of 1 corresponds to use in couplings between one and
two-photon states.

We now consider the solution of the problem in the case of a basis
with no more than two photons.
Eq.~(\ref{eq:e1}) provides the solution for the bare coupling between one
and two-photon states and, through spectator dependence of $\psi_0$,
makes $e_1$ a function of the final electron momentum.  We then need to
consider the bare coupling between the bare electron and the
one-photon states.
On substitution of Eqs.~(\ref{eq:CouplingRenormCond}) and
(\ref{eq:TproperFormula}), Eq.~(\ref{eq:e0}) becomes
\begin{equation} \label{eq:e0New}
e_0=\sqrt{Z_{2f}Z_{2i}}\frac{V_{fi}}{T_{fi}}e_R\,.
\end{equation}
This is a nonlinear equation for $e_0$ because $T_{fi}$ has
a complicated dependence on this bare charge.  To make this
dependence explicit, we first use the fact that $|\Psi\rangle$
is an eigenstate of $H_0+V$ to reduce (\ref{eq:TfiFormula}) to
the form
\begin{equation} \label{eq:TfiExplicit}
T_{fi}=\left[m_e^2-\frac{m_e^2+\pi^2/L_\perp^2}{1-2/K}
               -\frac{m_\gamma^2+\pi^2/L_\perp^2}{2/K}\right]
               \psi_0\psi_1^*(1-2/K,\hat{x}\pi/L_\perp)\,.
\end{equation}
The amplitude $\psi_1$ satisfies the middle equation of
(\ref{eq:IntegralEqns}), which can be written as
\begin{equation} \label{eq:psi1eq}
\bbox{b}_1\psi_0 + [\bbox{A}'+e_0^2\bbox{c}\bbox{c}^\dagger]\cdot\bbox{\psi}_1
     =m_e^2\bbox{\psi}_1\,,
\end{equation}
where $\bbox{A}'$ is an effective interaction obtained by integrating out
the $\psi_2$ amplitude and the $e_0^2\bbox{c}\bbox{c}^\dagger$ term
is the finite instantaneous fermion interaction.  The coupling to
$\psi_0$ contains $e_0$ as a simple factor.  We extract this in
the following definitions:
\begin{equation}
\bbox{b}_1^\prime\equiv\bbox{b}_1/e_0\;\; \mbox{and}\;\;
\bbox{\psi}_1^\prime\equiv\bbox{\psi}_1/(e_0\psi_0)\,.
\end{equation}
The scaled amplitude $\psi_1^\prime$ can then be obtained
as a formal solution to Eq.~(\ref{eq:psi1eq}) that shows
all $e_0$ dependence\footnote{Notice that $\bbox{A}'$ is independent
of $e_0$, as can be seen from (\ref{eq:Psi1s}) and (\ref{eq:Psi2s}).}
\begin{equation} \label{eq:psi1Prime}
\bbox{\psi}_1^\prime=\bbox{B}
    -e_0^2\frac{\bbox{c}^\dagger\cdot\bbox{B}}
              {1+e_0^2\bbox{c}^\dagger\cdot\bbox{D}}\bbox{D}\,,
\end{equation}
with
\begin{equation}
\bbox{B}\equiv-(\bbox{A}'-m_e^2)^{-1}\bbox{b}_1^\prime\;\;\mbox{and}\;\;
\bbox{D}\equiv (\bbox{A}'-m_e^2)^{-1}\bbox{c}\,.
\end{equation}
The amplitude for two-photon states is then given by
\begin{equation}
\bbox{\psi}_2=e_0\psi_0\bbox{\psi}_2^\prime=e_0\psi_0 R\bbox{\psi}_1^\prime\,,
\end{equation}
where $R$ is a rectangular matrix independent of $e_0$.

The discrete normalization condition is
\begin{equation}
1=|\psi_0|^2+|\bbox{\psi}_1|^2+|\bbox{\psi}_2|^2
 =|\psi_0|^2[1+e_0^2|\bbox{\psi}_1^\prime|^2+e_0^2|\bbox{\psi}_2^\prime|^2]\,.
\end{equation}
This yields
\begin{equation}
\psi_0=1/\sqrt{1+e_0^2(|\bbox{\psi}_1^\prime|^2+|\bbox{\psi}_2^\prime|^2)}\,,
\end{equation}
which can be used with (\ref{eq:TfiExplicit}) and (\ref{eq:e0New})
to obtain
\begin{equation}
e_0=\frac{\sqrt{Z_{2i}}V_{fi}^\prime e_R
         \sqrt{1+e_0^2(|\bbox{\psi}_1^\prime|^2+|\bbox{\psi}_2^\prime|^2)}}
      {\psi_1^{\prime *}(1-2/K,\hat{x}\pi/L_\perp)
        \left[m_e^2-\frac{m_e^2+\pi^2/L_\perp^2}{1-2/K}
               -\frac{m_\gamma^2+\pi^2/L_\perp^2}{2/K}\right]}\,,
\end{equation}
where $V'\equiv V/e_0$ is independent of $e_0$.
The phases of $\psi'_1$ and $V'_{fi}$ are such that the
right-hand side is real, as it must be.
The remaining implicit dependence on $e_0$ is in $\bbox{\psi}_1^\prime$,
which is given by (\ref{eq:psi1Prime}),
and in $\bbox{\psi}_2^\prime=R\bbox{\psi}_1^\prime$, with $R$ independent of
$e_0$.  Notice that $\bbox{B}$ and $\bbox{D}$ are independent of $e_0$ and
need to be computed only once.
The equation for $e_0$ is best solved iteratively after it is squared
to eliminate the square root on the right hand side.

\begin{figure}
\centerline{\epsfxsize=\columnwidth \epsfbox{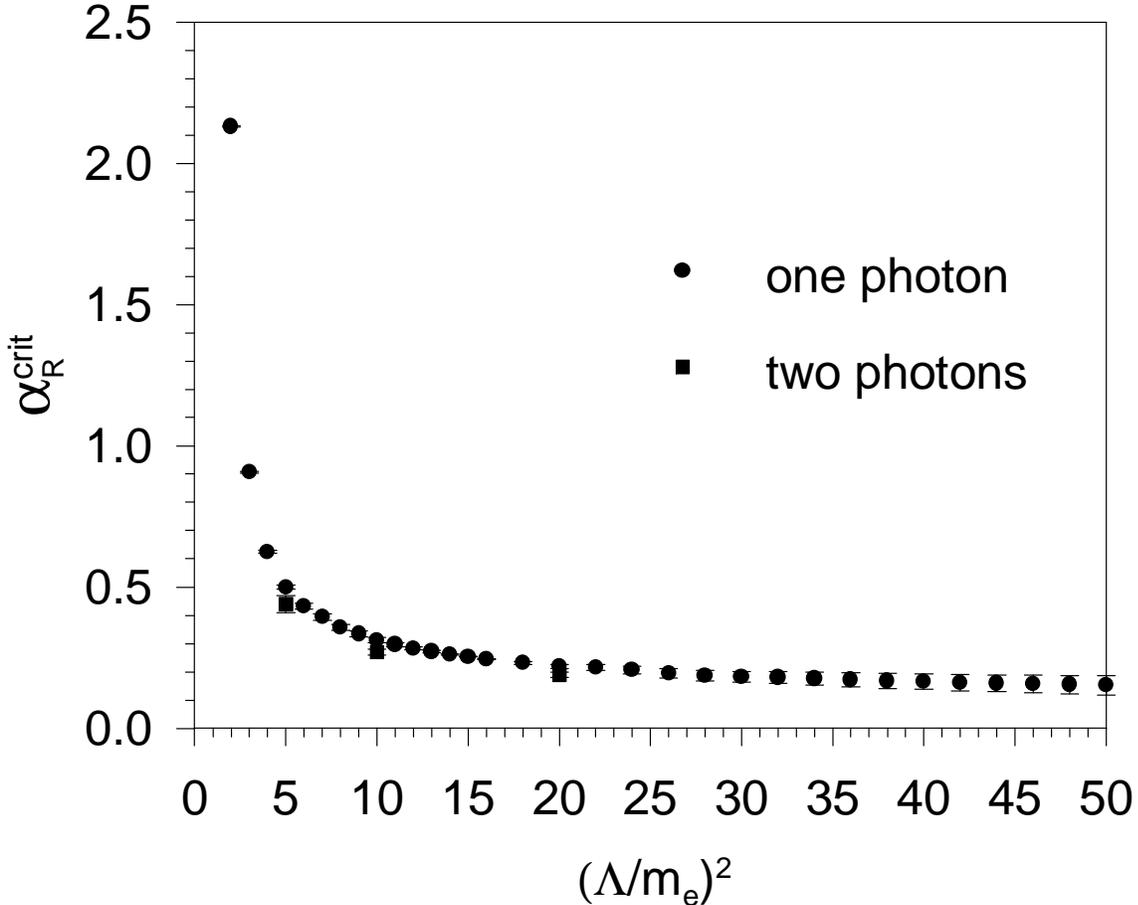} }
\caption{\label{fig:CritCoupling}
The critical coupling $\alpha_R^{\rm crit}$ as a function of cutoff $\Lambda$.
This is the upper limit of the range allowed for the physical coupling;
beyond this value there is no solution for the bare coupling,
as a consequence of the truncations of the theory.
The one-photon values are obtained from extrapolations using
$\alpha_R^{\rm crit}+a_1/K+a_2/K^2+b_1/N_\perp+b_2/N_\perp^2+c_{11}/(KN_\perp)$;
the error bars represent the difference between this fit and one without the
$b_1/N_\perp$ term.  The two-photon values come from a fit to
$\alpha_R^{\rm crit}+a_1/K+a_2/K^2+b_1/N_\perp+b_2/N_\perp^2$;
the error bars are obtained from a fit without the quadratic terms.  For
all the photon mass is $m_\gamma=m_e/10$.}
\end{figure}

A real solution exists only for a finite range of the physical coupling
$e_R$.  This is an artifact of the Tamm--Dancoff truncation and the
consequent failure of the Ward identity.  The value of the critical coupling
$e_R^{\rm crit}$, the upper limit of the allowed range,
can be found by studying the $e_0\rightarrow\infty$
limit.  In this limit we find
\begin{equation}
\alpha_R^{\rm crit}=\frac{(e_R^{\rm crit})^2}{4\pi}
    =\frac{\left[\psi_1^{\prime *}(1-2/K,\hat{x}\pi/L_\perp)\right]^2}
    {4\pi Z_{2i}V_{fi}^{\prime\,2}
        (|\bbox{\psi}_1^\prime|^2+|\bbox{\psi}_2^\prime|^2)}
    \left[m_e^2-\frac{m_e^2+\pi^2/L_\perp^2}{1-2/K}
               -\frac{m_\gamma^2+\pi^2/L_\perp^2}{2/K}\right]^2\,,
\end{equation}
with $\bbox{\psi}'_1$ calculable as
$\bbox{B}-(\bbox{c}^\dagger\cdot\bbox{B}
                      /\bbox{c}^\dagger\cdot\bbox{D})\bbox{D}$.
Values of the critical coupling are plotted in Fig.~\ref{fig:CritCoupling}.
The change from a basis with no more than one photon to a basis with no
more than two is quite small.  To stay within the limit imposed by
this result we will use a physical coupling of $\alpha_R=0.1$.

\subsection{Infrared singularities}   \label{sec:Infrared}

A nonzero photon mass $m_\gamma$ is used to eliminate the usual
infrared singularities.  Because the calculation deals with a
charged system, there would otherwise be considerable difficulty
with soft photons\cite{Misra}.  On the light cone, there are other
singularities not removed by the photon mass.  They are associated with
contributions that involve zero longitudinal momentum.

\begin{figure}
\centerline{\epsfxsize=\columnwidth \epsfbox{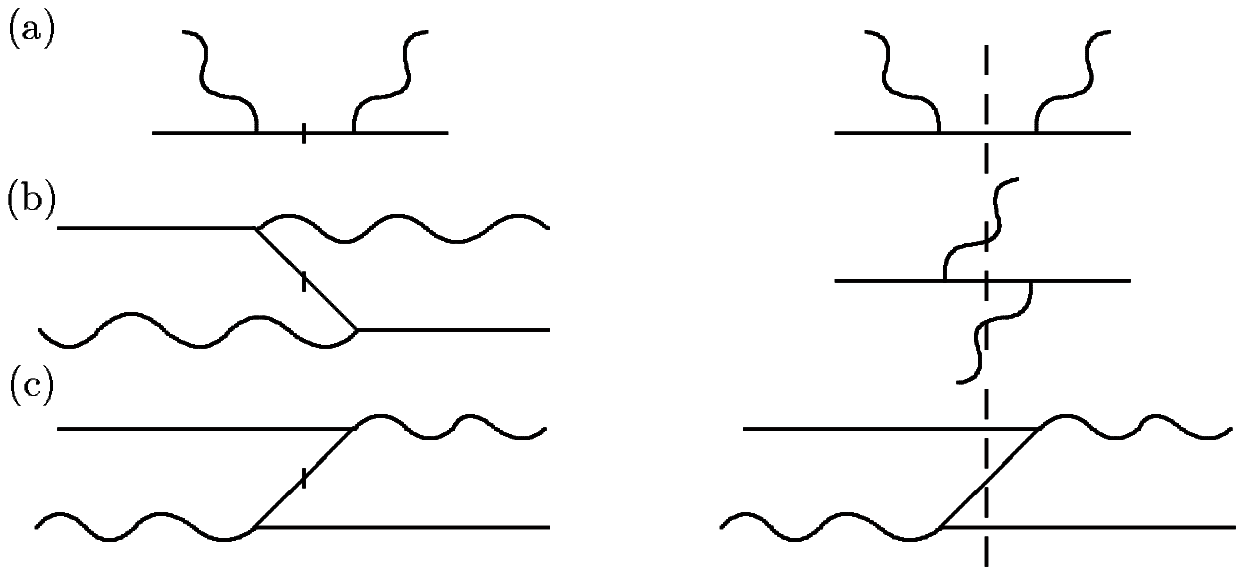} }
\caption{\label{fig:InfraredCancel}
Instantaneous fermion interactions paired with corresponding
iterated three-point interactions.  The vertical dashed lines indicate
the intermediate states of the iterated interactions.  In (b) the
longitudinal momentum of the instantaneous fermion is positive, and
in (c) it is negative; this represents a separation of the
crossed-photon graph into two pieces.}
\end{figure}

The fundamental four-point vertices can be infrared singular, in the
limit of zero longitudinal momentum for the instantaneous fermion.
They must be allowed
to cancel against iterations of the three-point vertices which are also
singular.  This constrains the bare couplings in the four-point vertices
to forms derived from the three-point coupling $e_0$.  The pairs of diagrams
are shown in Fig.~\ref{fig:InfraredCancel}.  The first pair does not actually
involve a singularity; however, we do match the four-point coupling to
the iterated three-point coupling.
The second requires basis states with two photons, which are
available in the calculation.  The third
pair requires the presence of electron-positron pairs in the basis,
or an effective interaction in the Hamiltonian.
Neither is included at present, and therefore this piece of the
instantaneous interaction must also be excluded from the
Hamiltonian.
Detection of $k^+<0$ in the instantaneous
interaction can be easily done to exclude this graph.

Other infrared singularities are associated with the emission and
absorption of real photons with longitudinal momentum near
zero\cite{Casher}.  In perturbative calculations these are regulated
by the Mandelstam--Leibbrandt prescription\cite{MLprescrip}.
Viewed in $x^+$-ordered perturbation theory, each intermediate state
contributes a denominator in which the light-cone energy
$k^-=(m_\gamma^2+k_\perp^2)/k^+$
of the photon becomes large, and each vertex can contain a factor of
$(k^+)^{-3/2}$.  If $k^+$ is separately regulated, with some lower
cutoff $\epsilon$, graphs with multiple photons will contribute powers
of $\log\epsilon$ or even $\epsilon^{-1}$.  For a Tamm-Dancoff approximation
to a charged system, these
cannot be expected to cancel.  The choice of the invariant mass
cutoff (\ref{eq:cutoff}) instead couples the regulation
of small $k^+$ to that of large $k_\perp$.
The combination prevents the small $k^+$ region of integration
from making large contributions except in cases where there are already
ultraviolet transverse divergences.  These spurious infrared
infinities are then handled by the mass and coupling renormalization
discussed in this section.

\subsection{Four-point graphs}  \label{sec:FourPtGraphs}

\begin{figure}
\centerline{\epsfxsize=\columnwidth \epsfbox{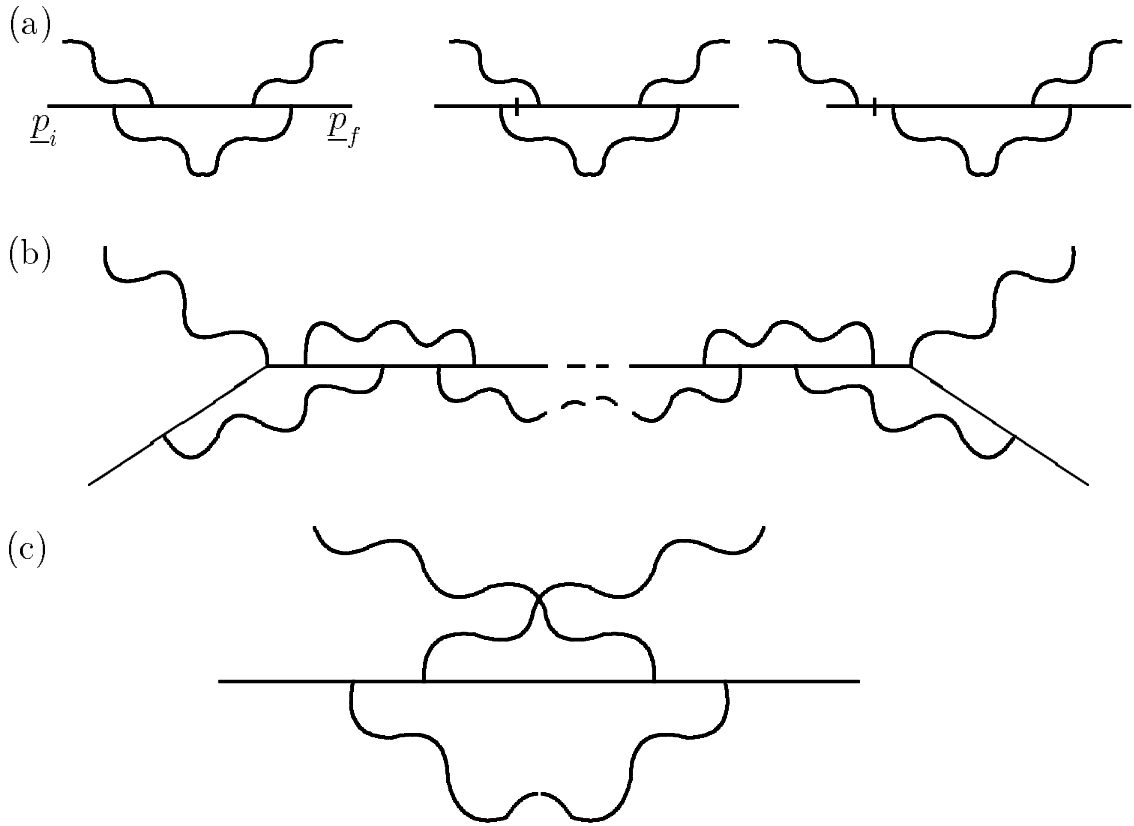} }
\caption{\label{fig:FourPointGraphs}
Logarithmically divergent four-point graphs.  Those with no more than
two photons in any intermediate state, (a) and (b), are included in the
present calculation.  The nonperturbative nature of the calculation
implies that (b) can have an arbitrary number of loops.  Diagram (c),
which contains three photons in flight is not included.}
\end{figure}

There remains a logarithmic divergence associated with four-point
graphs of the sort illustrated in
Fig.~\ref{fig:FourPointGraphs}(a) and (b)\@.
If all graphs of this order are included in a perturbative calculation,
the logarithms cancel.  However, the Tamm-Dancoff truncation of the
present calculation excludes some graphs,
such as the one shown in Fig.~\ref{fig:FourPointGraphs}(c), and the
cancellation can no longer take place.  In a nonperturbative calculation
one must include the equivalent of diagrams with an arbitrary number of
interlocked loops, such as Fig.~\ref{fig:FourPointGraphs}(b), which are
also logarithmically divergent.  The needed counterterm is of the form
$\lambda(p_i^+,p_f^+)\log\Lambda$ but cannot be found analytically without
summing all the infinite chains of interlocked loops.

One way to approach the construction of $\lambda(p_i^+,p_f^+)$ is to fit
Compton amplitudes to data\cite{Compton}.  This will require development
of techniques to describe scattering processes within DLCQ.
A generalization of earlier work\cite{Ree,Haydock}
on the inversion of the full Greens function may be useful
as a means for computing the $T$ matrix and thus
scattering amplitudes.  We do not consider this further here.
The results presented in Sec.~\ref{sec:Results} do still contain
the $\log\Lambda$ divergence.

\section{Finite corrections} \label{sec:FiniteCorrections}

\subsection{Photon zero modes}

As applied to QED, DLCQ requires the use of periodic boundary conditions
for the photon field.  This is because photons couple to fermion
bilinears, which are automatically periodic, even if the preferred
antiperiodic boundary conditions are used for the fermions.  For fields
periodic in the longitudinal direction $x^-$, there are contributions
from zero modes\cite{Vary,QEDzeromodes,ZeroModes,SSB,Phi4}, modes independent
of $x^-$ that correspond to zero longitudinal momentum.  As shown by Pauli and
Kalloniatis\cite{QEDzeromodes}, these modes prevent the choice of
ordinary light-cone gauge because the zero-mode piece of $A^+$ cannot
be gauged away.  Instead this piece must satisfy a constraint equation.
In fact, careful application of DLCQ to most bosonic theories will result
in constraint equations that relate the zero-mode contribution to the
normal-mode operators in a nonlinear, nontrivial way.  For QED there are
zero modes in all the components of the photon field, for which constraint
equations must be solved.  What is more, the constraint equation for
the dependent piece of the fermion field, which is easily solved in
light-cone gauge in the continuum, becomes coupled to the zero-mode
constraint equations.  A formulation of the coupled system of
constraints has been given by Kalloniatis and
Robertson\cite{KalloniatisRobertson}.  Extension of these constraints
to include a nonzero photon mass is straightforward.

The constraint equations are difficult to solve, even in simpler
theories\cite{Phi4}.  This is partly because they couple states with
different $P^+$ and require study of convergence as a function of
some $P^+$ cutoff.  The difficulty is also due to the need for an
ultraviolet cutoff and renormalization of masses and couplings.
Because the renormalization is formulated in terms of solutions
to the mass eigenvalue problem and because the Hamiltonian cannot be formed
until the zero-mode contribution is known, the problem expands to
a very large nonlinear system of simultaneous equations.  As a result
of these difficulties, the calculations discussed here do not include
zero modes.  However, some progress has been made
recently by Kalloniatis\cite{Kalloniatis} in the solution of
constraint equations for SU(2) Yang-Mills in $1+1$ dimensions 
coupled to massive adjoint scalars.

For theories such as QED where symmetry breaking effects are not expected,
solution of constraint equations may not be necessary. One can instead treat
the end-point behavior of photon amplitudes in a manner similar to that of the
``ladder relations'' studied by Antonuccio and Dalley\cite{LadderRelations}.
Behavior of amplitudes at small longitudinal momentum, as
extracted from the integral equations,
can be used to construct effective interactions that include
zero modes to leading order in $1/K$.  This is equivalent to
the approach used in\cite{Wivoda} where the behavior of the exchange
kernel was studied in a scalar theory to determine the effective
interaction\cite{RobertsonMaeno}.
To keep zero-mode terms to higher order in $1/K$ would actually
be inconsistent with DLCQ's neglect of higher order non-zero-mode terms.
In the work of Ref.~\cite{Wivoda} inclusion of the zero-mode
contributions ${\cal O}(1/K)$ did improve convergence.

The whole issue may actually be moot when the invariant mass cutoff
(\ref{eq:cutoff}) is used.  This cutoff explicitly excludes
contributions from states
with zero longitudinal momentum.  The meaning of this exclusion for
nondynamical fields is unclear.  The calculations that showed zero
modes to be useful for convergence\cite{Wivoda} did not employ the invariant
mass cutoff.  New calculations need to be carried
out specifically to study the effect of cutoff choice on the importance
of zero modes.

\subsection{Restoration of symmetries}

The use of light-cone coordinates, combined with the Tamm-Dancoff truncation
in particle number and the invariant mass cutoff, explicitly break symmetries
of the theory\cite{SectorDependent}.  In particular, rotational
symmetry about the transverse axes is broken because the associated
operators involve the interaction and therefore change particle number.
The change in particle number cannot be accommodated in field theory
without allowing an infinite number of particles.

Restoration of such symmetries can be accomplished by the addition
of finite counter-terms to the
Hamiltonian\cite{BurkardtLangnau,CovariantCurrent} including adjustment
of the ``vertex mass,'' which appears in the spin-flip vertex,
relative to the ``kinetic mass''\cite{BurkardtLangnau}.
The ambiguities associated with the infinite counterterms
allow such finite terms
to exist\cite{SectorDependent}.  Restoration of symmetries is then viewed as
a source of conditions by which these finite parts can be determined.  In
practice, this might involve study of processes\cite{Burkardt} such as Compton
scattering\cite{Compton} or electron-electron scattering.

Given the Tamm-Dancoff truncation, an alternative is to view the eigenvalue
problem as a few-body problem\cite{AngCond} for which the correct effective
Hamiltonian and the generators of translations, rotations, and boosts must
satisfy the usual Poincar\'{e} algebra\cite{PoincareAlgebra}.  The effective
operators might be constructed by adding minimal finite corrections to
their field-theoretic forms.  The finite corrections are determined by
the requirement that the Poincar\'{e} algebra be satisfied.

For the results presented here, no attempt has been made to include
these finite corrections.

\section{Results}  \label{sec:Results}

\begin{figure}
\centerline{\epsfxsize=\columnwidth \epsfbox{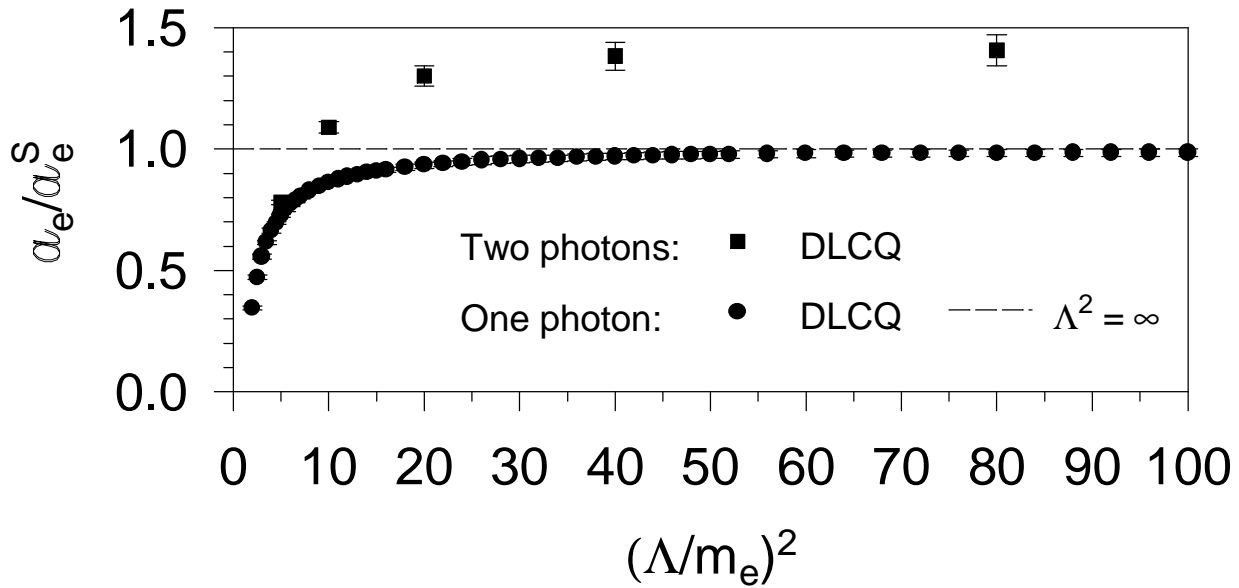} }
\caption{\label{fig:Results}
The ratio of the renormalized anomalous moment $a_e$ to
the exact one-photon perturbative result $a_e^S$
at the same photon mass $m_\gamma$,
as a function of the invariant-mass cutoff $\Lambda$.
Here $m_\gamma=m_e/10$, and $\alpha_R=0.1$.  The Fock space is
truncated to include at most one or two photons.
The DLCQ results are extrapolated from calculations done with
$K=21$ to 31, and $N_\perp=8$ to 15 for one photon and
$N_\perp=7$ to 9, 10, or 11 for two photons.}
\end{figure}

An accurate DLCQ calculation for
a basis with at most one photon can be easily done when instantaneous
interactions are neglected.  The accuracy
can be verified directly because the integrals that yield $a_e$
can be performed analytically\cite{BrodskyDrell}. In the limit
of infinite cutoff and zero photon mass this reproduces the
Schwinger\cite{Schwinger}
result of $\alpha_R/2\pi$.  The only coupling renormalization is
a trivial wave function renormalization.  The
DLCQ result at various cutoff values is shown in 
Fig.~\ref{fig:Results} for a photon mass of $m_e/10$.
Weighting methods\cite{PV} are a critical part of the calculation.

Calculations with a basis that includes at most
two photons have been done at five different values of
the cutoff for a coupling of $\alpha_R=0.1$.
They are also shown in Fig.~\ref{fig:Results}.
The two-photon contribution adds approximately 40\%.
This is much larger than the order of magnitude
($\alpha/\pi$ or 3\%) that one would expect.
It is also opposite in sign to the
Sommerfield--Petermann contribution\cite{SommerfieldPetermann}
of $-0.328(\alpha/\pi)^2$ to the anomalous moment.
We attribute this large difference to the absence of
Z graphs.

The basis sizes involved are on the order of 1 to 4
million, which translates to solution of linear systems with
4000 to 10,000 variables once the two-photon states are 
integrated out and symmetries of the one-photon states are 
used.

The values obtained from DLCQ were extrapolated to $K=\infty$ and
$N_\perp=\infty$ by fits to $a_e+a/K+b/N_\perp+c/(KN_\perp)$.
Exclusion of the last term provided an estimate of the error in
the fit, which is reflected in the error bars in Fig.~\ref{fig:Results}.
The values of $K$ ranged from 21 to 31 and those of $N_\perp$ from
7 to 9, 10 or 11, depending on matrix size limitations.

The time required for an
extrapolated value at a fixed cutoff is roughly
10 hours on a Cray X-MP, using less than 32 million words of
memory.  This seems quite competitive with older lattice
methods, where a quenched QCD calculation of heavy-light meson
wave functions\cite{HeavyLight} required 300 hours on a
CM-200\cite{JunkoPrivComm}, but does not yet match the effort
required with the latest methods\cite{Lepage}, for which
calculation of the B meson magnetic form factor might require
50 hours on a good PC\cite{LepagePrivComm}.

\section{Summary} \label{sec:Summary}

The nonperturbative calculation of the
anomalous moment of the electron $a_e$, besides being of intrinsic interest
itself,  exposes many important issues for nonperturbative
calculations within gauge theories which occur in the context of a
truncated Fock space.  These include nonperturbative mass and coupling
renormalization, control of spurious infrared singularities, determination
of zero-mode contributions, and
the construction of finite counterterms which restore symmetries. Each of these
has been addressed in the preceding sections, and the first two have been
incorporated into a DLCQ calculation where as many as two photons are included
in the basis.

We have presented results for $a_e$ computed in a light-cone-gauge
Fock space truncated to include one bare electron and at most two photons;
{\em i.e.},
up to two photons in flight.  The calculational scheme uses an invariant mass
cutoff, discretized light-cone quantization (DLCQ), a Tamm--Dancoff truncation
of the Fock space, and a photon mass regulator.  We have utilized new weighting
methods which greatly improve convergence to the continuum within DLCQ.
A large renormalized coupling strength $\alpha_R= 0.1$ is then used
to make the nonperturbative effects in the electron anomalous moment from the
one-electron, two-photon Fock state sector numerically detectable.
Results are given in Fig.~\ref{fig:Results}.

The disagreement between
these results and what one would expect from perturbation theory at
order $\alpha^2$ indicates that the effect of Z graphs needs to be included in a
systematic way.  This can be done as an effective interaction, to avoid
expansion of the Fock basis to include pair states.  The corresponding piece of
the instantaneous fermion interaction, as depicted in
Fig.~\ref{fig:InfraredCancel}(c), must then also be included to maintain an
infrared cancellation.

Further progress in
computing the electron moment will require:

1. New counterterms: One piece of the infinite renormalization is missing in
the calculation. As discussed in Section~\ref{sec:FourPtGraphs}, it requires a
new nonperturbative counterterm for the logarithmic divergences present in
diagrams of the type shown in Fig.~\ref{fig:FourPointGraphs}(a) and (b).  The
divergence arises because the Tamm--Dancoff truncation prevents certain
cancellations.  Construction of the counterterm will likely require analysis of
scattering processes.

2. Zero modes in DLCQ: Before full consideration of photon zero modes is
undertaken, we recommend renewed study of zero modes in a scalar theory
where the constraint equation can be solved exactly\cite{Wivoda}. This may show
that, when the invariant mass cutoff (\ref{eq:cutoff}) is used, zero modes do
not make a significant numerical contribution. If instead there is an important
contribution, it should be computed only to leading order in the numerical
resolution, to be consistent with the level of approximation used in the
basic DLCQ approach.

3. Use of symmetries in DLCQ renormalization: The restoration of symmetries
should then complete construction of the light-cone Hamiltonian.  One can
normalize to specific physical processes\cite{Burkardt,Compton} or take an
abstract approach based on the algebra of the Poincar\'{e}
generators\cite{AngCond,PoincareAlgebra}.

4. Higher Fock States: Once the two-photon calculation is
fully under control, the addition of $eee^+$ states can be considered.  This
will require analysis of photon mass and wave function renormalization.

Many of the complications of the light-cone Fock state analysis presented here
can be traced to the complexity of sector-dependent renormalization.
Given such complications the newly developed
alternative of Pauli--Villars regularization\cite{PV} may be the preferred
approach.  Within such a scheme, it is also likely that the limitation to a
small number of photons can be relaxed.

The analysis presented here is the first step in a systematic program to compute
physical  quantities in gauge theory systematically utilizing a light-cone
Fock expansion.
It will also be interesting to use these methods and the present knowledge of
the dressed-electron state in QED in order to systematically construct the
neutral positronium state as a composite of a dressed electron and positron.
Such an analysis can serve  as
the prototype for systemic nonperturbative construction of colorless bound
state hadrons in QCD.

\acknowledgments

This work was supported in part
by the Minnesota Supercomputer Institute through grants of computing time
and by the Department of Energy.  We thank G. McCartor, R. Perry,
St.\ G{\l}azek, D. Robertson, and A. Kalloniatis for helpful discussions.

\end{document}